\documentclass[twocolumn]{aastex631}
\usepackage{natbib}
\usepackage{amsmath}
\usepackage{gensymb}
\begin{document}
\newcommand{\Teff}{T_\mathrm{eff}}
\newcommand{\logg}{\log{g}}
\title{Evidence for the Late Arrival of Hot Jupiters in Systems with High Host-star Obliquities}
\author[0000-0002-7993-4214]{Jacob H. Hamer}
\affiliation{William H. Miller III Department of Physics and Astronomy, Johns Hopkins University, 3400 N. Charles Street, Baltimore, MD 21218, USA}
\correspondingauthor{Jacob H. Hamer}\email{jhamer3@jhu.edu}

\author[0000-0001-5761-6779]{Kevin C. Schlaufman}

\affiliation{William H. Miller III Department of Physics and Astronomy, Johns Hopkins University, 3400 N. Charles Street, Baltimore, MD 21218, USA}
%\email{kschlaufman@jhu.edu}

\received{2021 December 20} \revised{2022 March 29} \accepted{2022 April 21} \submitjournal{\aj} %\published{2019 October 21}
\begin{abstract}
\noindent
It has been shown that hot Jupiters systems with massive, hot stellar
primaries exhibit a wide range of stellar obliquities.  On the other
hand, hot Jupiter systems with low-mass, cool primaries often have
stellar obliquities close to zero.  Efficient tidal interactions between
hot Jupiters and the convective envelopes present in lower-mass main
sequence stars have been a popular explanation for these observations.
If this explanation is accurate, then aligned systems should be older
than misaligned systems.  Likewise, the convective envelope mass of a hot
Jupiter's host star should be an effective predictor of its obliquity.
We derive homogeneous stellar parameters---including convective
envelope masses---for hot Jupiter host stars with high-quality
sky-projected obliquity inferences.  Using a thin disk stellar
population's Galactic velocity dispersion as a relative age proxy, we
find that hot Jupiter host stars with larger-than-median obliquities are
older than hot Jupiter host stars with smaller-than-median obliquities.
The relative age difference between the two populations is larger for
hot Jupiter host stars with smaller-than-median fractional convective
envelope masses and is significant at
the 3.6-$\sigma$ level.  We identify stellar mass, not convective envelope
mass, as the best predictor of stellar obliquity in hot Jupiter systems.
The best explanation for these observations is that many hot Jupiters
in misaligned systems arrived in the close proximity of their host stars
long after their parent protoplanetary disks dissipated.  The dependence
of observed age offset on convective envelope mass suggests
that tidal realignment contributes to the population of aligned hot
Jupiters orbiting stars with convective envelopes.

\end{abstract}


\section{Introduction}
Over the past quarter century, exoplanet discoveries have continually challenged the expectations set by our solar system. In the solar system, the total planetary angular momentum is aligned to within 7$^\circ$ of the rotational angular momentum of the Sun \citep{Beck2005}. This seemed to be a natural consequence of planet formation: the Sun and the protosolar nebula both inherited their angular momenta from the gas cloud that formed the solar system. Therefore the expectation was that exoplanets would also have angular momenta aligned with their host stars' rotation.

Through the Rossiter-McLaughlin (RM) effect, spectroscopic observations of a star during the transit of another body in the system can be used to infer the sky-projected stellar obliquity\footnote{Sky-projected stellar obliquities may also be inferred through other methods like starspot crossings \citep[e.g.,][]{Nutzman2011, SanchisOjeda2013}. Stellar obliquities projected along the line of sight can be inferred via the $v\sin{i}$ technique \citep[e.g.,][]{Schlaufman2010}, asteroseismology \citep[e.g.,][]{Huber2013}, or other related techniques. However, most hot Jupiter obliquity measurements are made via the RM technique.}, $\lambda$, of the eclipsed star \citep{Rossiter1924, McLaughlin1924}. The earliest applications of the RM effect to constrain stellar obliquities in hot Jupiter systems revealed well-aligned systems that conformed with the expectations set by the solar system \citep[e.g.,][]{Queloz2000, Winn2005, Winn2006}. But this conformity would not last - RM measurements of the XO-3 system showed for the first time that not all exoplanet systems have small projected stellar obliquities \citep{Hebrard2008, Winn2009}. Unlike the disk-driven migration scenario described by \citet{Lin1996}, other hot Jupiter formation mechanisms are consistent with misalignments. Instead the formation of hot Jupiters was explained via interactions with other planets or stars that raised their inclinations and eccentricities until tidal dissipation in the planet circularized the orbit at their present locations \citep[e.g.,][]{Fabrycky2007, Nagasawa2008}.

The measurement of sky-projected stellar obliquities for several tens of hot Jupiter systems revealed an apparent relationship between the stellar obliquities of hot Jupiters and the mass/temperature of their host stars. Hot Jupiters orbiting less massive, cooler stars tended to be well-aligned, while those orbiting higher mass, hotter stars exhibited a wider range of spin-orbit angles \citep{Schlaufman2010, Winn2010}. 

There are several possible ways that the pattern could be interpreted. It could be that the pattern is indicative of separate formation channels for hot Jupiters orbiting cooler and hotter stars. For example, it may be that the relationship between stellar mass and the occurrence rate of giant planets \citep[e.g.,][]{Bowler2010, Ghezzi2018} facilitates the formation of misaligned hot Jupiters via planet-planet scattering followed by high-eccentricity migration. However, it is difficult to explain how such a scenario could produce a sharp transition in the range of stellar obliquities as a function of host star mass.

It is also possible that the pattern is not unique to hot Jupiters, and therefore is unrelated to hot Jupiter formation or tidal evolution. Indeed, studies of stellar obliquities in Kepler-discovered systems with lower-mass, longer-period planets that experience negligible tidal interactions have uncovered a seemingly similar pattern \citep[e.g.,][]{Mazeh2015, Louden2021}.  But comparisons between the obliquity distribution of the host stars of planets discovered by Kepler and the obliquity distribution of hot Jupiter host stars have shown that hot Jupiter systems with both hot and cool primaries exhibit a wider range of stellar obliquities than the host stars of planets discovered by Kepler. The implication is that different processes are at play in the two populations \citep{Winn2017, Munoz2018}.  

It may be that all hot Jupiters share a common formation channel that produces misaligned systems. Tidal realignment that occurs quickly in systems with low-mass primaries and more slowly in systems with high-mass primaries could then produce the observed stellar obliquity pattern. One aspect of this tidal realignment hypothesis that was initially difficult to reconcile with tidal theory was the requirement that realignment should occur more quickly than the orbital decay of the planet \citep[e.g.][]{Barker2009}. But later theoretical work showed that the spin-orbit misalignment could activate modes of tidal dissipation whereby the realignment could take place without a requisite decay of the orbit \citep[e.g.][]{Lai2012, Lin2017, Anderson2021}. 

The transition in the range of observed obliquities was tied to the transition in stellar structure that occurs at approximately the same stellar mass \citep[e.g.,][]{Schlaufman2010, Winn2010, Albrecht2012, Dawson2014}. Above 1.2 $M_\odot$, where systems are more often observed to be misaligned, stars' surface convective envelopes disappear \citep[e.g.,][]{Pinsonneault2001}. This change in structure is thought to be very important, as tidal dissipation is expected to be more efficient in convective envelopes \citep[e.g.,][]{Zahn1977}. If the convective envelopes can decouple from the radiative cores, then the significant convective envelope masses expected in low-mass stars can facilitate the realignment of cool star photospheres \citep[e.g.,][]{Winn2010}. Additionally, in stars with convective envelopes and radiative cores, internal gravity waves may be able to produce efficient dissipation. In stars with this structure the waves generated at the radiative-convective boundary can propagate to the center of the star, where their energy is focused into a smaller volume, leading to wave-breaking that enhances the dissipation rate \citep[e.g.,][]{Goodman1998}. Alternatively or in addition, \citet{Dawson2014} attributed the observed relationship between stellar obliquity and mass to more efficient magnetic braking in lower-mass stars with convective envelopes. 

Several studies have put forward additional evidence that tidal realignment of stellar photospheres is responsible for the relationship between stellar mass and obliquity. \citet{Triaud2011} proposed a tentative correlation between stellar obliquity
and age for hot Jupiter systems with $M_{\ast} \ge 1.2\ M_{\odot}$.
He further suggested that this correlation provides evidence that
hot Jupiters form misaligned and early with realignment occurring
within 2.5 Gyr.  The small number of stellar ages available to \citet{Triaud2011} were heterogeneously inferred without the benefit of the precise
parallaxes now available, though. \citet{Albrecht2012} constructed a theoretical tidal realignment timescale parameter and used it to show that systems with shorter realignment timescales were well-aligned. On the other hand, those systems with relatively longer realignment timescales had a random distribution of stellar obliquities. However, their parametrization relied on a parametric relation between $\Teff$ and convective envelope masses, heterogeneous data, and only half the number of obliquity inferences available at present. Similarly, \citet{Rice2021} construct a tidal realignment timescale parameter and find that planets with longer timescales for alignment tend to be observed with larger misalignments. \citet{Rice2021} also showed that systems with long eccentricity damping timescales do not follow the same relationship between host star effective temperature and stellar obliquity, suggesting that in these systems dissipative processes have not had time to sculpt obliquities. This further suggests that the pattern is a product of tides, but they too relied on heterogeneous data and used previously calculated models which related stellar effective temperature to convective envelope mass. \citet{TejadaArevalo2021} found evidence that hot Jupiters tidally spin up their host stars, corroborating earlier findings \citep[e.g.,][]{Pont2009, Husnoo2012, Brown2014, Maxted2015, Penev2016}. Similarly, \citet{Reggiani2022} found evidence for angular momentum transfer in the WASP-77 A system. However, the timescales for realignment and tidal spin up may not be the same, so this is not conclusive evidence of tidal realignment.

The availability of precise Gaia astrometry enables both precise, homogeneous planet host stellar parameter inference as well as model-independent relative stellar age inferences, so a reevaluation of the tidal realignment scenario as the explanation for the relationship between hot Jupiter host star mass and obliquity is now possible. In this paper, we use all available data to calculate precise, homogeneous photospheric stellar parameters and convective envelope masses for hot Jupiter host stars with obliquity inferences. We use the Galactic velocity dispersion of a thin disk stellar population as a proxy for its age and find that the host stars of misaligned hot Jupiters are older than the hosts of aligned systems. This age offset is more significant in the subsample with host stars that have smaller-than-median fractional convective envelope masses. We show that convective envelope mass is not predictive of stellar obliquity and that host star mass or the scaled planet semimajor axis are the most meaningful predictors of obliquity. We suggest that misaligned hot Jupiter systems are formed after protoplanetary disk dissipation but more than a few hundred Myr before the present, and that tides lead to realignment in systems where the host star has a convective envelope. We describe the data we use in our analyses in Section 2. We detail in Section 3 the procedures we use to homogeneously derive photospheric stellar parameters and structures, to evaluate the relative ages of hot Jupiters with high and low sky-projected stellar obliquities $\lambda$, and to model the relationship between system parameters and $\lambda$. We discuss the possible explanations for our observation and its implications for our understanding of hot Jupiter formation in Section 4. We conclude in Section 5.

\section{Data}
We retrieve all known hot Jupiters from the NASA Exoplanet Archive \citep{Akeson2013} as of February 15, 2022 following the definition of \citealt{Wright2012}: $P<10\ $days and $M_\mathrm{pl}>0.1\ M_\mathrm{Jup}$. We apply this mass cut to the best mass that is available; we use masses or mass upper limits in order of descending priority.%\edit1{We use the radius cut to select hot Jupiters without mass measurements but whose radii imply a mass $M_\mathrm{pl}\gtrsim0.1\ M_\mathrm{Jup}$.}
~For the planets which only have mass upper limits, we remove those whose radii ($\lesssim0.8\ R_\mathrm{Jup}$; e.g., \citealt{Masuda2017}) do not securely imply a mass greater than 0.1 $M_\mathrm{Jup}$: Kepler-63 and TOI-942. We remove the brown dwarfs ($M_\mathrm{pl}>11 \ M_\mathrm{Jup}$) KELT-1, CoRoT-3, and HATS-70, as brown dwarfs likely form differently from planet-mass objects \citep[e.g.,][]{Schlaufman2018}. We use the default parameter set in all cases except K2-24 b and XO-3 b. As the K2-24 b default parameter set does not include a planet mass, we use the planet mass and orbital period described in \citet{LilloBox2016}. In the case of XO-3 b, the default parameter set gives a planet mass of $\sim 7\ M_\mathrm{Jup}$ derived using a mass for XO-3 of 0.58 $M_\odot$. But detailed analyses of XO-3 produce mass estimates closer to $\sim1.2\ M_\odot$, resulting in a planet minimum mass $\gtrsim11\ M_\mathrm{Jup}$ \citep[e.g.,][]{Winn2008, Worku2022}. Therefore, we treat XO-3 b as a brown dwarf and remove it from the sample. %We add HIP 67522 b to the sample, which does not have a mass measurement as its young host star is highly active, but is Jupiter-sized \citep{Heitzmann2021}.

We obtain sky-projected stellar obliquities for this sample by matching it with the TEPCat orbital obliquity catalog\footnote{\url{https://www.astro.keele.ac.uk/jkt/tepcat/obliquity.html}, queried February 15, 2022} compiled by \citet{Southworth2011}. We use all meaurements irrespective of the method for inferring $\lambda$. For each host, we use the measurement which is flagged as the best value. We use the true obliquity for systems which have had it measured. As we are only concerned with the absolute degree of misalignment, we take the absolute value $|\lambda|$. For systems which had $\lambda<0^\circ$, we flip their upper and lower uncertainties accordingly. A $\lambda$ greater than 180$^{\circ}$ is equivalent to $\lambda-360^{\circ}$ \citep[e.g.,][]{Benomar2014}. Therefore, for systems with $|\lambda|>180^{\circ}$ we reassign $|\lambda|$ as $|\lambda-360^{\circ}|$. 

We remove several systems from the sample due to possible systematic errors affecting their obliquity measurements \citep[e.g., Appendix C of][]{Albrecht2022}. We remove CoRoT-1 because the two measurements of its obliquity are strongly discrepant, with one favoring an aligned configuration \citep{Bouchy2008} and one favoring a misaligned configuration \citep{Pont2009}. We remove CoRoT-19 b \citep{Guenther2012} due to the lack of post-egress data and because the RM signal was detected at marginal significance. We remove HATS-14 b \citep{Zhou2015} due to the lack of post-egress data and because the retrieved obliquity is sensitive to the choice of $v\sin{i}$ prior. For similar reasons, we remove TOI-1268 b \citep{Dong2022} due to the lack of post-egress data and because the obliquity constraint is degenerate with the poorly-constrained impact parameter.

\subsection{Data for Galactic Velocity Dispersion Analysis}
The data queried from the NASA Exoplanet Archive include the Gaia DR2 \texttt{source\_id} for the hot Jupiter host stars. Next, we use the Gaia EDR3 \texttt{dr2\_neighbourhood} table provided by the Gaia Archive \citep{GaiaArchive} to obtain the Gaia EDR3 \texttt{source\_id} which may correspond to the DR2 \texttt{source\_id}. This table does not provide a one-to-one matching between the catalogs. We use only pairs of DR2/EDR3 ids which have a $G$ magnitude difference less than 0.1 mag. Following this requirement, there are no cases in which we must choose between multiple possible identifiers. Finally, we query the Gaia Archive to obtain all Gaia EDR3 data\footnote{For the details of
Gaia EDR3 and its data processing, see \citet{GaiaMission,GaiaEDR3}, \citet{Gaia3Catalogue}, \citet{Gaia3Astrometric}, \citet{Gaia3Photometry}, \citet{Gaia3PSF},
\citet{Gaia3RV}, \citet{Gaia3Source}, \citet{Gaia3Xmatch}.} for these EDR3 \texttt{source\_id}. While a common quality cut to ensure precise astrometry and remove spurious parallaxes is to require that $\texttt{RUWE}<1.4$, where $\texttt{RUWE}$ is the renormalised unit weight error, we do not apply this cut as some of the processes that are theorized to form hot Jupiters require stellar companions which would result in a high \texttt{RUWE} for their host stars. 
%For objects with $\texttt{RUWE}>1.4$, we ensure that the Gaia EDR3 parallaxes are not spurious measurements by checking that the Gaia DR2 and Gaia EDR3 parallaxes for the object are not discrepant at more than 3-$\sigma$. This requirement only results in the removal of one system, WASP-103. 
Nevertheless, we can be confident that the remaining sample still has precise astrometry, as the median $\texttt{parallax\_over\_error}$ in the sample is $\approx$228, with a minimum value of $\approx$17.  

In order to calculate Galactic space velocities, we need radial velocities in addition to the Gaia EDR3 astrometry. We use radial velocities (listed in order of decreasing priority) from the California-Kepler-Survey \citep[CKS;][]{Petigura2017}, the NASA Exoplanet Archive, the LAMOST DR7 Medium-Resolution Spectroscopic Survey \citep[MRS;][]{LAMOST, LAMOST2}, Gaia DR2 \citep{GaiaDR2RVs}, and the LAMOST DR7 Low-Resolution Spectroscopic Survey \citep[LRS;][]{LAMOST, LAMOST2}. In order to ensure that the Gaia DR2 radial velocities are high quality, we require $\texttt{dr2\_rv\_nb\_transits}>3$, following \citet{Marchetti2021}. We combine multiple LAMOST radial velocities for a given object using a weighted average. We weight each measurement by the radial velocity uncertainty for the MRS, and by the $g$-band SNR of the spectrum for the LRS. We remove LAMOST radial velocities with null uncertainties. The majority of the radial velocities are from Gaia DR2. The resulting sample of hot Jupiter hosts with the necessary data to calculate Galactic space velocities as well as sky-projected stellar obliquities contains 102 systems. The typical planet mass in the final sample is between 0.52 $M_\mathrm{Jup}$ and 3.37 $M_\mathrm{Jup}$, and the typical radius is between 1.07 $R_\mathrm{Jup}$ and 1.57 $R_\mathrm{Jup}$.

\subsection{Data for Inference of Stellar Properties}
We then obtain photometry for our sample of hot Jupiter hosts to enable us to fit MESA Isochrones \& Stellar Tracks (MIST) isochrones to our data \citep{MIST0,MIST1}. We query the Gaia Archive for the Gaia DR2 photometry \citep{GaiaPhotValidation} of our hosts\footnote{We prefer Gaia DR2 photometry over Gaia EDR3 photometry as our prior experience using Gaia DR2 photometry informs which passbands we choose to use as input for the isochrones.}. 

We use the Gaia crossmatch best neighbor tables to identify the corresponding AllWISE \citep{AllWISE, AllWISE2, AllWISEdoi, WISEdoi}, 2MASS \citep{2MASS, 2MASSV, 2MASSdoi}, SDSS DR13 \citep{SDSSDR13, SDSSIV, SDSS_telescope, SDSS_camera, SDSS_filters, SDSS_photometric_system}, and SkyMapper Data Release 2 \citep{Skymapper, Skymapper_doi} identifiers. We query GALEX \citep{Galex, GalexV} on VizieR using the Gaia EDR3 designations of our host stars. We describe in the Appendix the quality cuts applied to each set of photometry to ensure high quality photometry.

We obtain spectroscopic estimates of the photospheric stellar parameters for use in the likelihood calculation of isochrones. We use photospheric stellar parameters (in descending order of priority) from the \citet{Brewer2016} and \citet{Brewer2018} catalogs, the SWEET-Cat catalog \citep{Santos2013, Andreasen2017, Sousa2018, Sousa2021}, and the NASA Exoplanet Archive. We use the statistical uncertainties described in Table 6 of \citet{Brewer2016} when using photospheric stellar parameters derived in that analysis. We limit the SWEET-Cat catalog to only the measurements marked as being homogeneously derived. In systems missing uncertainties on the photospheric stellar parameters, we assume uncertainties corresponding to the 84th percentile of the uncertainties on that parameter for the other stars in the catalog: 0.04 in [Fe/H], 100 K in $\Teff$, and 0.1 in $\logg$.

We need to estimate the extinction to the host stars in our sample in order to produce accurate photospheric stellar parameters when fitting isochrones. We query the Bayestar 3D dust map \citep{Green2014, Green2019}, using \texttt{dustmaps} \citep{dustmaps}, to estimate the line-of-sight reddening to the host stars in our sample. We use the 50th percentile of the reddening samples at the distance and position queried as the best value, and define the uncertainty to be the mean of the difference between the 84th and 50th percentiles and the difference between the 50th and 16th percentiles.
We convert the reddening estimate of the map to an $A_V$ extinction by multiplying by 2.742, the coefficient for Landolt V in Table 6 of \citet{Schlafly2011}\footnote{\url{http://argonaut.skymaps.info/usage}}.
% We convert the reddening estimate of the map to an E($g-r$) using the coefficients in \citet{Green2019} Table 1. We then convert this to an E$(B-V)$ reddening by multiplying by 0.981.\footnote{\url{http://argonaut.skymaps.info/usage\#eq-1}} Finally, we use $A_V = 3.1\mathrm{E}(B-V)$ to obtain $A_V$.  
The reddening estimate is accompanied by two quality flags. The first flag indicates that the Markov Chain Monte Carlo fit to the line-of-sight reddening converged. The second flag tells the user if the distance being queried falls outside of the range of distances of the stars used to construct the differential reddening measurement. We do not use the reddening estimate when either of these two quality flags fail. In these cases, we supplement with the dust map described in \citet{Capitanio2017}. This map returns $\mathrm E(B-V)$, which is converted to an $A_V$ by multiplying by 3.1.

\section{Analysis}
\subsection{Inference of Host Star Properties}
We use the \texttt{isochrones} package to homogeneously derive stellar parameters for our sample of hot Jupiter hosts with measured sky-projected stellar obliquities. First, we initialize a single star model in \texttt{isochrones} using a MESA Isochrones \& Stellar Tracks (MIST) stellar model \citep{Paxton2011, Paxton2013, Paxton2015, Paxton2018, Paxton2019, Morton2015, MIST0, MIST1}. We pass this model the GALEX NUV, SkyMapper $uvgriz$ or SDSS $ugriz$, Gaia $G$, 2MASS $JHK$, and AllWISE $W1$ \& $W2$ magnitudes as available. If the $u$ band is available in both SkyMapper and SDSS, we use whichever survey has more bands with measurements. If the $u$ band is available in only one of the surveys, we use the survey for which it is available. If neither survey provides a $u$ band measurement, then we use whichever survey has more bands with measurements. If an object has no photometric data other than from Gaia, we also supply the Gaia $G_\mathrm{BP}$ and $G_\mathrm{RP}$ magnitudes. We only include these magnitudes in this case as our experience has shown that including these data in addition to other photometry reduces the overall quality of the fit.

We also supply the model with the estimated line-of-sight extinction, the Gaia EDR3 parallax, and the spectroscopic stellar parameters from \citet{Brewer2016}, \citet{Brewer2018}, or SWEET-Cat \citep[][]{Santos2013, Andreasen2017, Sousa2018, Sousa2021}, or the NASA Exoplanet Archive. Because the photospheric stellar parameters in the NASA Exoplanet Archive are heterogeneous, we only use them in our analyses when reddening-and metallicity-sensitive $u$ or GALEX $NUV$ photometry are unavailable. These bands are crucial for accurately constraining the extinction of our host stars, and therefore are crucial for accurately determining stellar parameters. Therefore, we include these heterogeneous data only when it is necessary as a prior on the stellar parameters. 

We use default priors on all parameters except extinction, distance, and age. We use a flat prior for distance between $1000/(\pi+5\sigma_\pi)$ and $1000/(\pi-5\sigma_\pi)$ where $\pi$ is the parallax in mas and $\sigma_\pi$ is its uncertainty. As all of our systems are nearby and have precise astrometry, the error introduced by inverting parallax to obtain these distances is negligible. For systems with $A_V+3e_{A_V}<0.6$ according to the dust maps, we use a power law prior with an index of -0.5 and bounded between 0.00001 and $\max{(0.15, A_V+3e_{A_V})}$, where $e_{A_V}$ is the uncertainty on the dust map extinction estimate. This prior is motivated by the small distance to our host stars, and is approximately fit to match the distribution of extinction values from the dust maps. For the remaining systems, we use a flat prior between 0 and $\max{(1, A_V+e_{A_V})}$. We use a flat prior in $\log_{10}$ stellar age between ($10^8$, $11\times10^9$) years. This only differs from the default age prior in that it excludes ages between $10^7$ and $10^8$ years in order to avoid degeneracies between young and old ages for several stars which had no indication of an age below 100 Myr. We then fit MIST isochrones to these data using MultiNest \citep{Feroz2008, Feroz2009, Feroz2019}. We use the median value of the derived samples for each stellar parameter as the best value, and use the 16th and 84th percentiles of the samples as lower and upper uncertainties. We visually inspect the output to ensure that all of the fits successfully converge. That is, we check that the fits do not converge against the edge of any prior and that there are no multimodalities with large discrepancies between the photospheric stellar parameters corresponding to each mode. 

\begin{figure*}[h]
    \centering
    \plotone{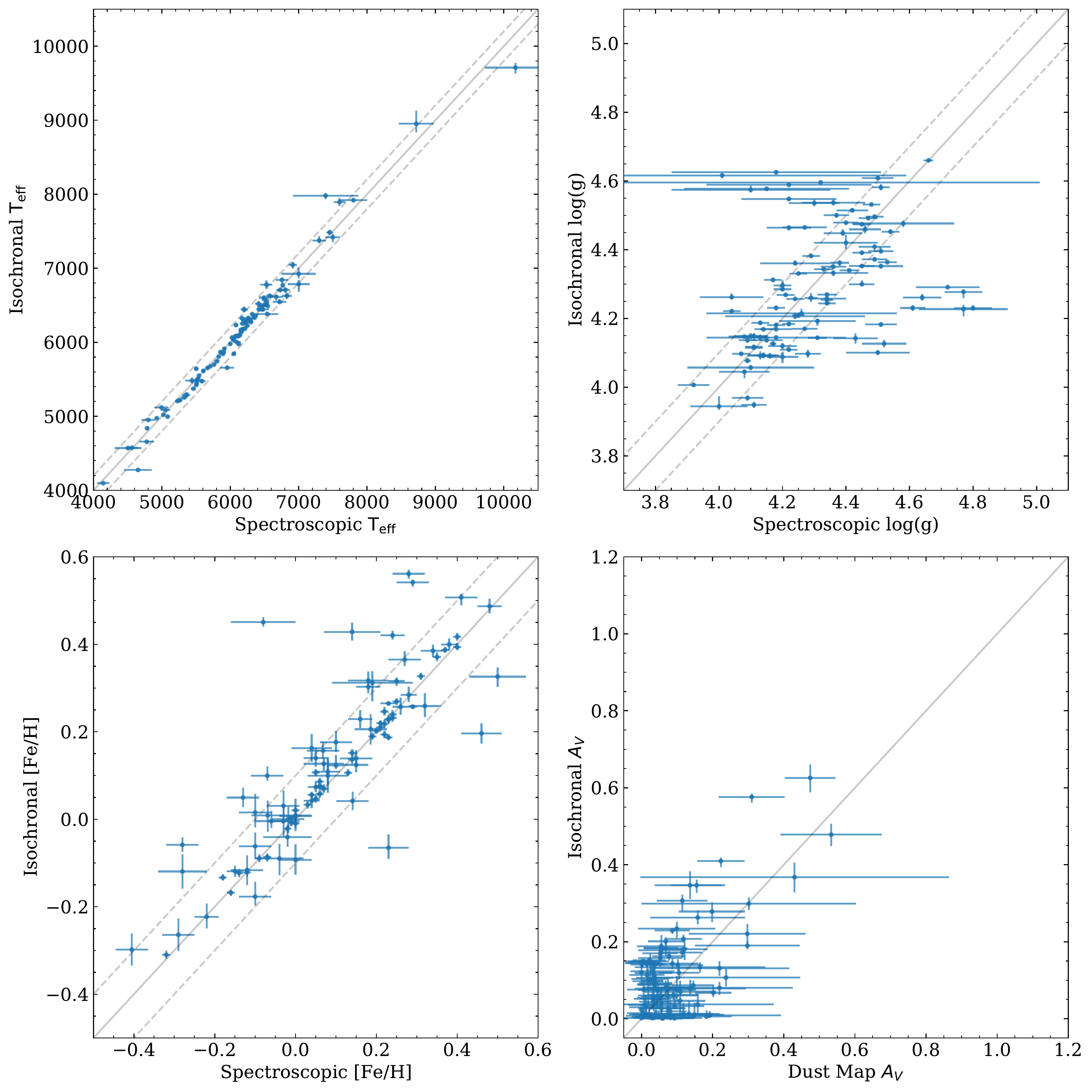}
    \caption{Comparison between the photospheric stellar parameters derived from high-resolution spectroscopy and isochrone analysis. Solid grey lines indicate the line $y=x$. Top left: Comparison between the stellar effective temperature predictions. Dashed grey lines indicate 200 K above and below the one-to-one correspondence. The parameters output by the two methods display good agreement. Top right: Comparison between the predicted surface gravities. The isochronal surface gravities are consistent with the surface gravities expected for main sequence stars. Dashed grey lines indicate 0.1 dex above and below one-to-one correspondence. Bottom left: Comparison between the metallicity estimates. Dashed grey lines indicate 0.1 dex above and below one-to-one correspondence. While in many cases the isochronal metallicity is higher, this is not unexpected as hot Jupiter host stars are preferentially metal-rich. Bottom right: Comparison between the extinction predictions of the 3D dust maps and those of isochrones. Most stars have low extinction, consistent with their small distances.}
    \label{fig:Figure 1}
\end{figure*}

We show in Figure~\ref{fig:Figure 1} a comparison between the spectroscopic estimates of the photospheric stellar parameters and the estimates output by isochrones. The isochronal and spectroscopic effective temperature estimates agree well. As the convective envelope mass of main sequence stars has the strongest relationship with effective temperature, this means that the convective envelope masses we derive should be of good quality. While some of the spectroscopic surface gravities are lower than our isochronal estimates, these spectroscopic values often have large uncertainties. Moreover, many of the more precise, underestimated spectroscopic surface gravities come from the Brewer catalogs, suggesting a systematic offset. We investigated the systems with isochronal surface gravities smaller than the spectroscopically-determined values and found that SWEET-Cat surface gravities are increasingly overestimated relative to isochronal values as effective temperature increases. The discrepant measurements likely reflect a systematic offset. Overall, as our isochronal modeling recovers spectroscopic effective temperatures well and is informed by precise and accurate Gaia parallaxes, our surface gravity estimates should be reliable. While there are some significant discrepancies between the isochronal and spectroscopic [Fe/H] estimates, the isochronal [Fe/H] tend to be higher, which is to be expected given that hot Jupiter hosts are preferentially metal rich. We find that SWEET-Cat metallicities tend to be underestimated more significantly overall and moreso for the cooler stars in the sample. As we constrain the extinction using dust maps and use GALEX NUV and SkyMapper or SDSS $u$-band photometry as available, our metallicities should be reliable. The most important quality of our photospheric stellar parameter estimates is their homogeneity, which is crucial for revealing any relationship between system parameters and obliquity such that the analysis is unaffected by systematic offsets between different parameter estimates.
 
We then use our homogeneously derived stellar parameters as input for stellar models, which allow us to estimate the convective envelope masses of our host stars. We use Modules for Experiments in Stellar Astrophysics (MESA, \citealt{Paxton2011, Paxton2013, Paxton2015, Paxton2018, Paxton2019}) to generate our stellar models. We base the necessary inlist files off of those that are part of the \texttt{1M\_pre\_ms\_to\_wd} test suite. The stellar model is initialized using the stellar mass and metallicity determined using isochrones. The total metal content of the star is defined as $0.0142\times10^{[\mathrm{Fe}/\mathrm{H}]}$, where [Fe/H] is the stellar metallicity with respect to the Sun \citep{Asplund2009}. We set \texttt{use\_Type2\_opacities} to False as these opacities are only relevant for stars which have evolved beyond the main sequence. Our sample is limited to main sequence stars, as reflected by their surface gravities. We terminate the model when it reaches the stellar age estimated by isochrones. We make the inlist files and output files for each star available on the JHU Data Archive\footnote{\url{https://doi.org/10.7281/T1/6I0KPP}}.

We then use the MESA output to calculate the fraction of stellar mass in the convective envelope of our host stars. We show in Table 1 the input data for the isochrone analysis. In Table 2 we report the data output by the isochrone analysis and the fractional convective envelope masses derived in the stellar modeling. 

\begin{figure}[h]
    \centering    
    \epsscale{1.3}
    \plotone{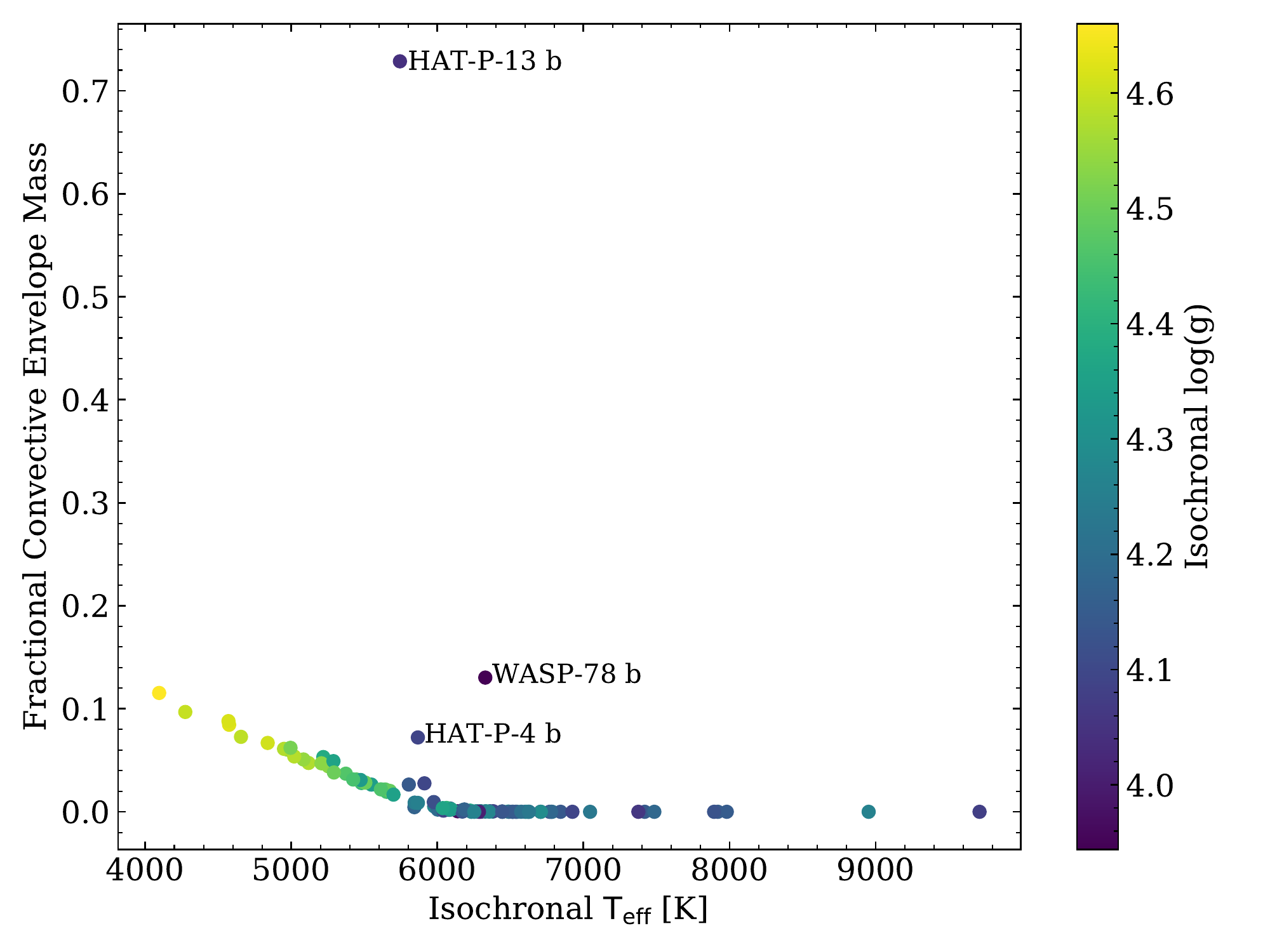}
    \caption{Fraction of stellar mass in the surface convective envelope as a function of $\Teff$. Point color indicates the $\logg$ predicted by isochrones. The output of the MESA models reproduces the expected general trend of increasing mass in the convective envelope at lower $\Teff$, with slightly evolved stars having more mass in their convective envelopes.}
    \label{fig:Figure 2}
\end{figure}

\begin{figure*}[h]
    \centering
    \epsscale{0.9}
    \plotone{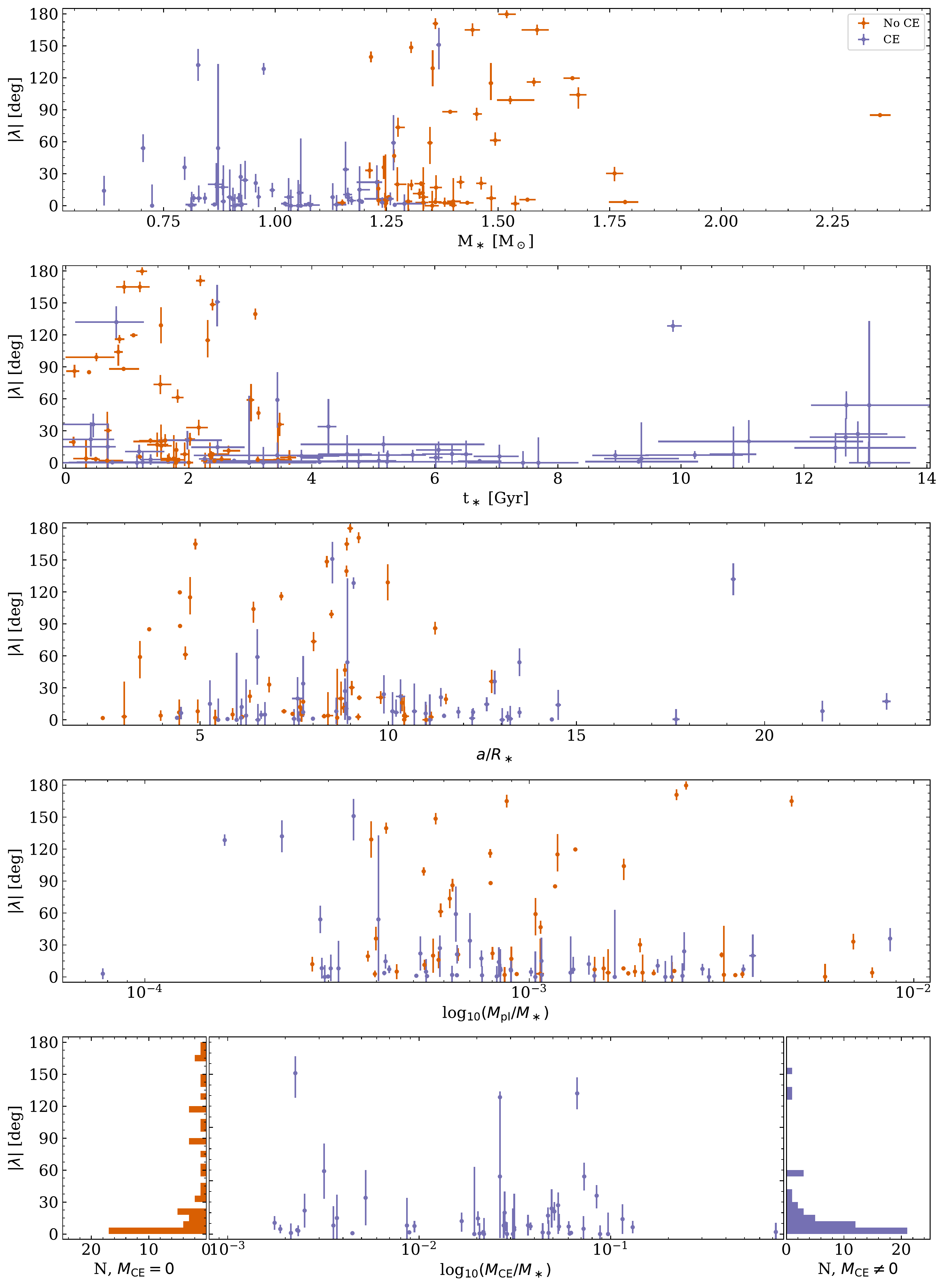}
    \caption{Sky-projected stellar obliquities of transiting hot Jupiter host stars plotted against stellar mass, stellar age, $a/R_\ast$, planet-to-star mass ratio, and convective envelope mass as a fraction of stellar mass. Blue points indicate systems for which our modeling of the host star determines it has a surface convective envelope. Red points indicate systems for which the host star has no mass in a surface convective envelope. According to tidal theory, systems with small scaled semi-major axes and larger planet-to-star mass ratios should align more quickly. According to the \citet{Albrecht2012} model for realignment, systems with more mass in the host star's convective envelope should align more quickly. Top: Stellar mass, $M_{\ast}$. $M_{\ast}$ alone is not a perfect proxy for the convective envelope mass, motivating our use of stellar models. Top-middle: Stellar age, $t_\ast$. Older systems appear to more often have low stellar obliquities, but stellar mass confounds any apparent relationship. Middle: Scaled semimajor axis $a/R_{\ast}$. Misalignments persist even at small values of $a/R_{\ast}$ in contrast to the strong $a/R_{\ast}$ dependence of tides. Bottom-middle: Mass ratio $M_{\text{p}}/M_{\ast}$.  Higher mass-ratio systems appear to be more well-aligned. Bottom-left: Distribution of stellar obliquities for systems in which the host star has no convective envelope. Bottom-middle: Fraction of stellar mass in the convective envelope, $M_\mathrm{CE}/M_\ast$.  High obliquities are present regardless of the presence of the convective envelope. Bottom-right: Distribution of stellar obliquities for systems in which the host stars has mass in their convective envelopes.}
    \label{fig:Figure 3}
\end{figure*}

We plot in Figure~\ref{fig:Figure 2} the fraction of the stellar mass in the surface convective envelopes of our stars against the stars' effective temperatures and surface gravities. The resulting convective envelope masses show the expected dependence on stellar mass and surface gravity. In other words, convective envelope mass increases as effective temperature decreases, and slightly evolved stars with lower surface gravities have more mass in their convective envelopes. 
We show in Figure~\ref{fig:Figure 3} the sky-projected stellar obliquities of the systems in our sample plotted against system parameters that may be relevant in determining the tidal realignment timescale. The system parameters were calculated using the stellar parameters we homogeneously derived. 

\begin{splitdeluxetable*}{cCCCCCCCBCCCCCCCBCCCCCC}
\tabletypesize{\footnotesize}
\tablecaption{Isochrone Analysis Input Data}
\tablenum{1}
\tablehead{\colhead{Host} &\colhead{Parallax} & \colhead{$A_V$} & \colhead{$T_\mathrm{eff,spec}$} & \colhead{$\log{g}_{\mathrm{spec}}$} & \colhead{[Fe/H]$_\mathrm{spec}$} & \colhead{NUV} & \colhead{$u$} & \colhead{$v$} & \colhead{$g$} & \colhead{$r$} & \colhead{$i$} & \colhead{$z$} & \colhead{$G_\mathrm{BP}$} & \colhead{$G_\mathrm{G}$} & \colhead{$G_\mathrm{RP}$} & \colhead{$J$} & \colhead{$H$} & \colhead{$K$} & \colhead{$W1$} & \colhead{$W2$}\\
\nocolhead{Host} &\colhead{(mas)} & \nocolhead{$A_V$} & \colhead{(K)} & \nocolhead{$\logg$} & \nocolhead{[Fe/H]} & \colhead{(mag)} & \colhead{(mag)} & \colhead{(mag)} & \colhead{(mag)} & \colhead{(mag)} & \colhead{(mag)} & \colhead{(mag)} & \colhead{(mag)} & \colhead{(mag)} & \colhead{(mag)} & \colhead{(mag)} & \colhead{(mag)} & \colhead{(mag)} & \colhead{(mag)} & \colhead{(mag)}}
\startdata
       WASP-32 &  3.57 $\pm$ 0.04 & 0.08 $\pm$ 0.05 &  6180 $\pm$  30 & 4.51 $\pm$ 0.04 &  0.05 $\pm$ 0.02 & 16.22 $\pm$ 0.01 & 13.034 $\pm$ 0.006 & 12.664 $\pm$ 0.004 &  -  &  - &  -  &  -  & 11.710 $\pm$ 0.002 & 11.4091 $\pm$ 0.0003 & 10.974 $\pm$ 0.002 & 10.50 $\pm$ 0.02 & 10.25 $\pm$ 0.03 & 10.16 $\pm$ 0.02 & 10.14 $\pm$ 0.02 & 10.19 $\pm$ 0.02\\
       WASP-26 &  3.96 $\pm$ 0.02 & 0.04 $\pm$ 0.05 &  6000 $\pm$  20 & 4.34 $\pm$ 0.03 &  0.14 $\pm$ 0.01 & 16.28 $\pm$ 0.01 & 12.729 $\pm$ 0.008 & 12.3356 $\pm$ 0.0010 &  -  &  -  &  -  &  -  & 11.314 $\pm$ 0.002 & 10.9935 $\pm$ 0.0003 & 10.538 $\pm$ 0.002 & 10.02 $\pm$ 0.02 &  9.78 $\pm$ 0.02 &  9.69 $\pm$ 0.02 &  -  &  - \\
       WASP-1 &  2.61 $\pm$ 0.02 & 0.15 $\pm$ 0.06 &  6120 $\pm$  30 & 4.13 $\pm$ 0.03 &  0.23 $\pm$ 0.01 & 16.62 $\pm$ 0.02 & 13.19 $\pm$ 0.01 &  - & - &  - &  - & 11.36 $\pm$ 0.02 & 11.844 $\pm$ 0.002 & 11.5349 $\pm$ 0.0003 & 11.085 $\pm$ 0.002 & 10.59 $\pm$ 0.02 & 10.36 $\pm$ 0.02 & 10.28 $\pm$ 0.02 & 10.22 $\pm$ 0.02 & 10.25 $\pm$ 0.02\\
       WASP-190 &  1.79 $\pm$ 0.02 & 0.02 $\pm$ 0.05 &  6730 $\pm$ 60 & 4.52 $\pm$ 0.07 &  0.15 $\pm$ 0.04 & 15.74 $\pm$ 0.01 & 13.135 $\pm$ 0.007 & 12.68 $\pm$ 0.01 &  -  & 11.635 $\pm$ 0.007 & 11.604 $\pm$ 0.007 &  - & 11.865 $\pm$ 0.002 & 11.6333 $\pm$ 0.0003 & 11.262 $\pm$ 0.002 & 10.86 $\pm$ 0.02 & 10.69 $\pm$ 0.02 & 10.65 $\pm$ 0.02 & 10.59 $\pm$ 0.02 & 10.62 $\pm$ 0.02\\
       HAT-P-16 &  4.50 $\pm$ 0.02 & 0.10 $\pm$ 0.07 &  6250 $\pm$  20 & 4.49 $\pm$ 0.04 &  0.21 $\pm$ 0.01 & 15.75 $\pm$ 0.01 &  -  & -  & -  & -  & -  & - & 11.053 $\pm$ 0.002 & 10.7568 $\pm$ 0.0003 & 10.327 $\pm$ 0.002 &  9.85 $\pm$ 0.02 &  9.62 $\pm$ 0.02 &  9.55 $\pm$ 0.02 &  9.51 $\pm$ 0.02 &  9.53 $\pm$ 0.02\\
\enddata
\tablecomments{Table 1 is published in its entirety and with full precision in the machine-readable format. A portion of the data is shown here with truncated precision and with the systems ordered by increasing right ascension.}
\end{splitdeluxetable*}
\begin{deluxetable*}{cCCCCCCCC}
\tabletypesize{\footnotesize}
\tablecaption{Isochrone Analysis and Stellar Modeling Output Data}
\tablenum{2}
\tablehead{\colhead{Host} & \colhead{$M_{\ast\mathrm{,iso}}$ } & \colhead{$R_{\ast\mathrm{,iso}}$} & \colhead{$\log_{10}{t_{\ast\mathrm{,iso}}}$} & \colhead{[Fe/H]$_\mathrm{iso}$} & \colhead{$T_{\mathrm{eff,iso}}$} & \colhead{$\logg_\mathrm{,iso}$} & \colhead{$A_{V\mathrm{,iso}}$} & \colhead{$M_\mathrm{CE}/M_\ast$}\\
\nocolhead{Host} & \colhead{($M_\odot$)} & \colhead{($R_\odot$)} & \nocolhead{$\log_{10}{t_{\ast\mathrm{,iso}}}$} & \nocolhead{[Fe/H]$_\mathrm{iso}$} & \colhead{(K)} & \nocolhead{$\logg_\mathrm{,iso}$} & \nocolhead{$A_{V\mathrm{,iso}}$} & \nocolhead{$M_\mathrm{CE}/M_\ast$}}
\startdata
       WASP-32 & $1.16_{-0.01}^{+0.01}$ & $1.13_{-0.01}^{+0.01}$ &  $9.08_{-0.18}^{+0.08}$ &  $0.07_{-0.01}^{+0.01}$ & $6180_{-20}^{+20}& 4.40_{-0.01}^{+0.01}$ & $0.16_{-0.01}^{+0.01}$ & 0.002 \\
       WASP-26 & $1.16_{-0.01}^{+0.01}$ & $1.31_{-0.01}^{+0.01}$ & $9.63_{-0.01}^{+0.02}$ & $0.15_{-0.01}^{+0.01}$ & $5980_{-10}^{+10}$ & $4.27_{-0.01}^{+0.01}$ & $0.09_{-0.02}^{+0.02}$ & 0.005 \\
       WASP-1 & $1.27_{-0.01}^{+0.01}$ & $1.51_{-0.01}^{+0.01}$ & $9.54_{-0.01}^{+0.01}$ & $0.19_{-0.01}^{+0.01}$ & $6087_{-6}^{+7}$ & $4.19_{-0.01}^{+0.01}$ & $0.09_{-0.01}^{+0.01}$ & 0.003 \\
       WASP-190 & $1.46_{-0.01}^{+0.01}$ & $1.73_{-0.02}^{+0.02}$ & $9.21_{-0.03}^{+0.02}$ &  $0.14_{-0.02}^{+0.02}$ & $6710_{-20}^{+20}$ & $4.13_{-0.01}^{+0.01}$ & $0.15_{-0.01}^{+0.01}$ & 0.000 \\
       HAT-P-16 & $1.25_{-0.01}^{+0.01}$ & $1.21_{-0.01}^{+0.01}$ &  $8.83_{-0.21}^{+0.08}$ &  $0.22_{-0.01}^{+0.01}$ & $6220_{-10}^{+10}$ & $4.37_{-0.01}^{+0.01}$ & $0.14_{-0.02}^{+0.01}$ & 0.001 \\
\enddata
\tablecomments{Table 2 is published in its entirety and with full precision in the machine-readable format. A portion of the data is shown here with truncated precision and with the systems ordered by increasing right ascension. The displayed uncertainties only represent the statistical uncertainties, and the systematic uncertainties are likely larger in most cases.}
\end{deluxetable*}
\subsection{Inferring Relative Ages with Galactic Velocity Dispersions}
We use a Monte Carlo simulation to determine the value of $|\lambda|$ that will typically divide the sample into aligned and misaligned halves. On each iteration of the simulation, we sample $|\lambda|$ from an asymmetric normal distribution representing its uncertainties for each system, and take $|\lambda|$ if it was scattered to a negative value or $|\lambda-360^\circ|$ if it was scattered to a value greater than $180^\circ$. At each iteration, we record the median value of $|\lambda|$ which divides the sample in half. The median value of the median across 1000 iterations is $12.8^\circ$. We use this value to define misalignment throughout the rest of our experiments.

Next, we use the kinematics of our hot Jupiter host stars to search for evidence that tides are acting to realign the photospheres of host stars of misaligned hot Jupiters. If the model of tidal realignment in which it is mediated by the convective envelope is valid and a single process generates hot Jupiters with a wide distribution of obliquities, then hot Jupiters orbiting stars with little-to-no mass in their convective envelopes should experience slow or negligible realignment. Aligned systems should either be older than misaligned systems or there should be no relative age difference between the populations. For hot Jupiters hosted by stars with more mass in their convective envelopes, the realignment process should be faster, and aligned systems should be older than misaligned systems. Because the velocity dispersion of a thin disk stellar population grows with the average age of that stellar population \citep[e.g.,][]{Binney2000}, we would expect that a sample of aligned hot Jupiters orbiting stars with little-to-no mass in their convective envelopes should have a significantly higher Galactic velocity dispersion than a sample of misaligned hot Jupiters orbiting stars with little-to-no mass in their convective envelopes, or that the two populations should have consistent Galactic velocity dispersion distributions. Similarly, we would expect that a sample of stars with more massive convective envelopes hosting aligned hot Jupiters should have a significantly higher Galactic velocity dispersion than a sample of stars with more massive convective envelopes hosting misaligned hot Jupiters.

We convert the astrometry and radial velocities of our host stars to Galactic space velocities.
For each star, we use \texttt{pyia} to generate 100 random samples from the Gaia uncertainty distribution, and then use \texttt{astropy} to calculate 100 realizations of the Galactocentric velocity \citep{PriceWhelan2018, astropy:2013, astropy:2018}. We calculate the Galactic velocity dispersion for each realization of the sample velocities as 
\begin{equation}
    \frac{1}{N}\sum \left[(U_i-\overline{U})^2+(V_i-\overline{V})^2+(W_i-\overline{W})^2\right]^{1/2},
\end{equation}
where $U$, $V$, and $W$ are the components of the Galactocentric velocity, and barred values are medians. We have 100 realizations of the Galactic velocity dispersion for the sample of stars at the end of this process. Finally, we use the median Galactic velocity dispersion across the 100 realizations as the Galactic velocity dispersion of that sample. 

We use a Monte Carlo experiment to account for the uncertainty on the sky-projected stellar obliquity while comparing the Galactic velocity dispersion of host stars of misaligned and aligned hot Jupiters. In a single iteration, we begin by sampling the sky-projected stellar obliquity from an asymmetric normal distribution representing its uncertainties. If $\lambda$ is scattered to a negative value, we take the absolute value. If $\lambda$ is scattered to be greater than $180^\circ$, we take $|\lambda-360^\circ|$. We then divide this sample into systems in which the host star has a fractional convective mass less than or more than the median fractional convective envelope mass in the sample. The median fractional convective envelope mass in the sample is 0.0015. In \citet{Albrecht2012}, ``hot" stars are defined as having $T_\mathrm{eff}>6250$ K because convective envelope mass drops below 0.002 $M_\odot$ at this effective temperature. This corresponds to a fractional convective envelope mass of 0.0017, close to our median value. We then further subdivide the samples into those which are aligned ($|\lambda|<12.8^\circ$)\footnote{We repeat the experiment using a cutoff of $21^\circ$ to test the sensitivity of the results to a larger cutoff value. We choose this value as it represents 3-$\sigma$ on the typical uncertainty on $\lambda$ in the overall sample. The results do not qualitatively change.}, and those which are misaligned ($|\lambda|>12.8^\circ$). For each of the four subsamples, we generate the Galactic velocity dispersion as described above and store its value. We repeat this process over 10000 iterations. 
\begin{figure*}[h]
    \centering
    \plotone{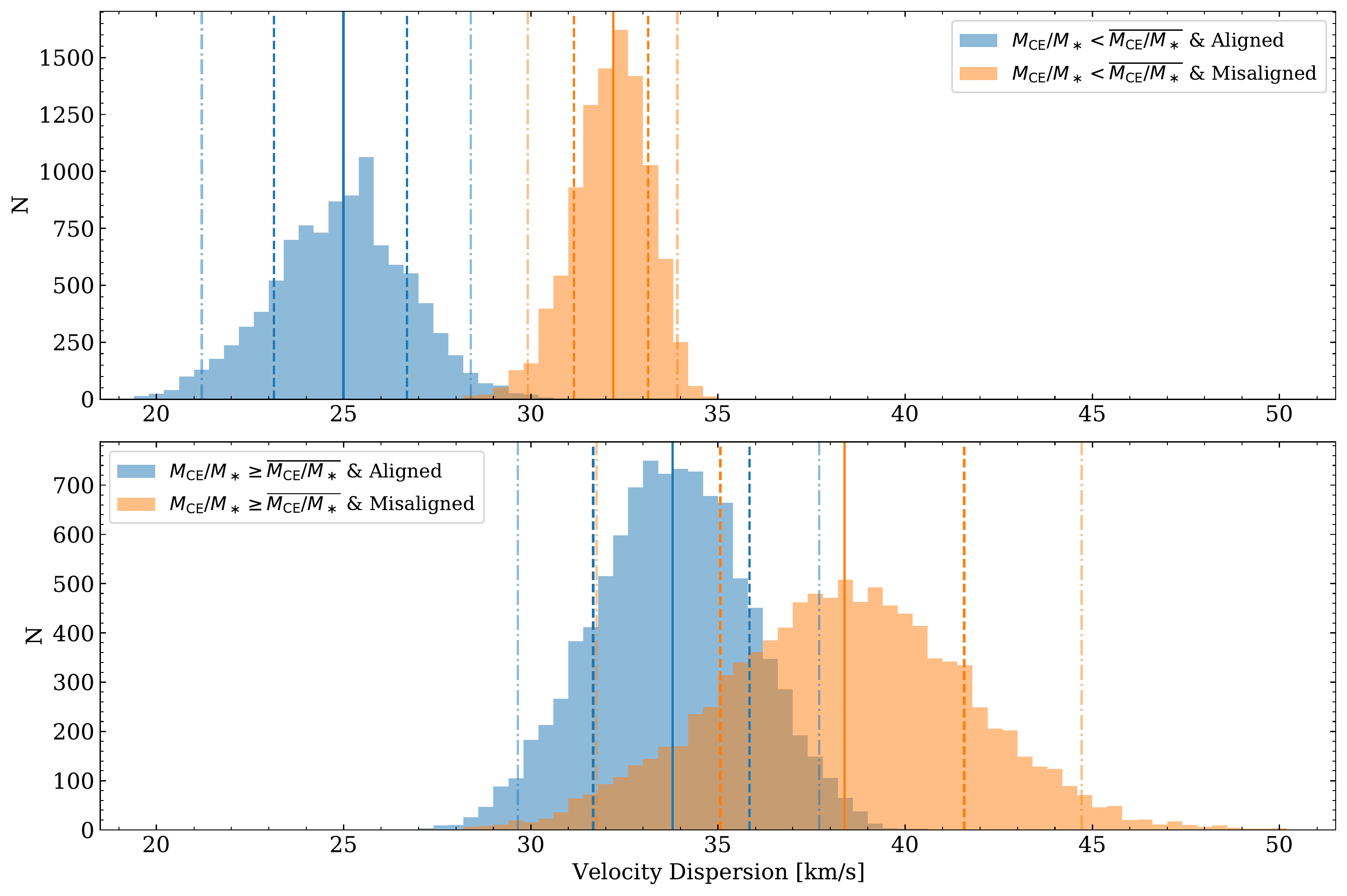}
    \caption{Distribution of Galactic velocity dispersion for Monte Carlo samples of hosts of aligned and misaligned hot Jupiters. At each iteration, the sky-projected stellar obliquity is sampled from within its uncertainty. Systems with $|\lambda|<12.8^\circ$ are considered aligned. The solid vertical lines indicate the median Galactic velocity dispersion of the samples, the dashed vertical lines indicate the 16th and 84th percentiles, and the dash-dot vertical lines indicate the 2.5th and 97.5th percentiles. Top: Systems for which the fractional convective envelope mass of the host star is less than the median value within the sample, and the alignment timescale should be long in the \citet{Albrecht2012} model. Across Monte Carlo realizations of $\lambda$ within its uncertainty, typical sample sizes for the misaligned and aligned systems are $31^{+2}_{-1}$ and $20^{+1}_{-2}$, respectively. We find that aligned systems have a smaller velocity dispersion than misaligned systems at roughly 3.6-$\sigma$, contrary to what would be expected if tides are damping misalignments over time. Bottom: Systems for which the host star has a fractional convective envelope mass larger than the median value within the sample, and the alignment timescale should be short. Typical sample sizes for the misaligned and aligned systems are $20^{+2}_{-3}$ and $31^{+3}_{-2}$, respectively. Aligned systems have a smaller velocity dispersion than misaligned systems at roughly 1.2-$\sigma$.}
    \label{fig:Figure 4}
\end{figure*}
We show the result of this Monte Carlo experiment in Figure~\ref{fig:Figure 4}. If tidal realignment is realigning the photospheres of stars hosting misaligned hot Jupiters, then aligned systems would have a larger Galactic velocity dispersion than misaligned systems. The result of this calculation is inconsistent with this prediction. For stars with smaller-than-median fractional convective envelope masses, hosts stars of misaligned hot Jupiters have a larger Galactic velocity dispersion than host stars of aligned hot Jupiters at 3.6-$\sigma$. For stars with larger-than-median fractional convective envelope masses, the same Galactic velocity dispersion offset is present but at only 1.2-$\sigma$. 

We also carry out an alternative version of this experiment in which we divide the misaligned and aligned subsamples in a different manner. As we had done earlier, we divide the sample in two based on the median fractional convective envelope mass. We generate Monte Carlo realizations of the sky-projected stellar obliquities. We calculate the median value of $|\lambda|$ within the two halves of the sample for each realization. In this construction, the misaligned and aligned subsamples are forced to be of equal size on each iteration rather than being of equal size on average. Across the Monte Carlo iterations, the typical $|\lambda|$ dividing the misaligned and aligned systems orbiting stars with smaller-than-median (larger-than-median) fractional convective envelope masses is $\sim22^\circ$ ($\sim8^\circ$). The results do not qualitatively change when carrying out the calculation in this manner. For stars with smaller-than-median (larger-than-median) fractional convective envelope masses, hosts stars of misaligned hot Jupiters have a larger Galactic velocity dispersion than host stars of aligned hot Jupiters at 2.2-$\sigma$ (0.9-$\sigma$).

These observations imply that misaligned systems are older than aligned systems, which cannot be possible if aligned hot Jupiters have been produced via the tidal realignment of misaligned hot Jupiters. In Section 4, we will discuss the implications of our observation for our understanding of hot Jupiter formation and evolution.

\subsection{Modeling the Relationship Between System Parameters and Sky-Projected Stellar Obliquities}
We perform a number of tests to determine what sets of parameters can best predict the degree of misalignment, having prepared a set of hot Jupiter hosts with sky-projected stellar obliquity measurements and homogeneously-derived convective envelope masses, stellar radii, stellar masses, and stellar ages. For this portion of the analysis, we remove the hot Jupiters in the sample which only have upper limits on their masses rather than $M_\mathrm{pl}$ or $M_\mathrm{pl}\sin{i}$: KELT-19A b, KELT-20 b, KELT-21 b, XO-6 b, WASP-167 b, and TOI-1518 b -- leaving 96 systems in the sample. The system parameters which we test for predictive power are those which should be important based on tidal theory, and specifically in the model put forward in \citet{Albrecht2012}: stellar mass $M_\ast$, stellar age $t_\ast$, planet-to-star mass ratio $M_\mathrm{pl}/M_\ast$, scaled planet semi-major axis $a/R_\ast$, and convective envelope mass $M_\mathrm{CE}$. As the realignment timescale scales as $(a/R_\ast)^6$ and $(M_\mathrm{pl}/M_\ast)^2$ \citep[e.g.,][]{Albrecht2012}, we fit models against $\log_{10}(a/R_\ast)$ and $\log_{10}(M_\mathrm{pl}/M_\ast)$. We generate the power set of this set of parameters - all subsets of these parameters.

For each set of parameters from the power set, we evaluate the ability of these parameters to predict misalignment. We implement a Monte Carlo experiment for which each iteration follows a cross-validation scheme, an established method for model selection that simultaneously serves to avoid overfitting the data. That is, on each iteration we select 70\% of the sample to train a model, and then evaluate how well it fits the data using the remaining 30\% of the sample as a test set. For the training set, we sample the the system parameters from within their uncertainties. For parameters with asymmetric uncertainties, we sample from within an asymmetric normal distribution. If a positive-definite parameter comes out of the sampling with a negative value it is reassigned to be $10^{-5}$. 
In the linear regression case, we use \texttt{statsmodels} ordinary least squares regression to fit a linear model between the set of system parameters being investigated and the Monte Carlo realization of $|\lambda|$. In the logistic regression case, we use \texttt{statsmodels} to perform logistic regression between the set of independent variables and a discrete variable based whether or not the Monte Carlo realization of $|\lambda|$ is less than or more than $12.8^\circ$. Additionally, we repeat the logistic regression experiment using a cutoff of 24$^\circ$ to determine if the results are sensitive to a larger cutoff value. We choose this value as it represents 3-$\sigma$ on the typical uncerainty on $\lambda$ in the subsample used in the regression analysis.

We compute metrics to evaluate how well the linear and logistic regression models that were generated on the training set fit the real test data. For the linear regression model, we compute the sum of squared errors (SSE) as 
\begin{equation}
    \sum_i^N \left(\lambda_\mathrm{true}-\lambda_\mathrm{pred}\right)^2/\sigma_\lambda^2
\end{equation}
where $\lambda_\mathrm{true}$ is the observed stellar obliquity, $\lambda_\mathrm{pred}$ is the stellar obliquity predicted by the linear regression model, and $\sigma_\lambda$ is the uncertainty on the observed stellar obliquity -- using the upper or lower uncertainty depending on if the predicted $\lambda$ is greater than or less than the observed value. Lower values of the sum of squared error indicate that a model makes more accurate predictions of the obliquity. For the logistic regression models we compute the Matthews correlation coefficient (MCC), a metric that evaluates the quality of a binary classifier, using \texttt{sklearn} \citep{sklearn}. The Matthews correlation coefficient is closer to 1 for classifiers which make correct classifications, is close to -1 for classifiers which typically produce false negatives and false positives, and is close to 0 when the classifier performs no better than assigning classes randomly. It is preferable to other classification metrics such as the F1 score and accuracy as it only takes on high values when a classifier effectively reproduces both true positives and true negatives, and as it is effective even for the imbalanced data sets which may result from our cross-validation technique \citep[e.g.,][]{Chicco2020}. We use these metrics to produce a rank-ordering of the models, shown below in Table 3.

\begin{deluxetable*}{|c|c|c|c|c|c|c|c|}
\tablewidth{5cm}
%\tablecaption{$\mathrm{BIC - BIC_{min}}$}
\tablenum{3}
\tablecaption{Model Comparison}
\tablehead{\colhead{$t_\ast$} & \colhead{$\log_{10}(a/R_\ast)$} & \colhead{$M_\mathrm{CE}$} & \colhead{$\log_{10}(M_\mathrm{pl}/M_\ast)$} & \colhead{$M_\ast$} & \colhead{Linear Regression SSE} & \colhead{Logistic Regression MCC, 12.8$^\circ$} & \colhead{Logistic Regression MCC, 24$^\circ$}}
%% All data must appear between the \startdata and \enddata commands
\startdata
x&&&&x& $\mathbf{4.91\times10^{4}}$ & 0.03 & 0.05\\ \hline
x&x&&x&x& 5.53$\times10^{4}$ & 0.01 & 0.07\\ \hline
x&x&&&x& 5.61$\times10^{4}$ & 0.00 & $\mathbf{0.11}$\\ \hline
x&&x&&x& 5.71$\times10^{4}$ & $\mathbf{0.09}$ & 0.04\\ \hline
&x&&&x& 5.77$\times10^{4}$ & 0.01 & 0.09\\ \hline
&&x&&x& 5.81$\times10^{4}$ & 0.07 &0.03\\ \hline
x&&&x&x& 5.81$\times10^{4}$ & 0.02 &0.07\\ \hline
&x&&x&x& 5.97$\times10^{4}$ & 0.02 & 0.09\\ \hline
&&&&x& 6.17$\times10^{4}$ & 0.00 & 0.00 \\ \hline
x&x&x&&x& 6.22$\times10^{4}$ & 0.06 & 0.09\\ \hline
x&x&x&x&x& 6.28$\times10^{4}$ & 0.05 & 0.05\\ \hline
x&&x&x&x& 6.35$\times10^{4}$ & 0.04 &0.04\\ \hline
&x&x&&x& 6.45$\times10^{4}$ & 0.07 &0.10\\ \hline
&&&x&x& 6.46$\times10^{4}$ & 0.02 &0.08\\ \hline
x&&&x&& 6.51$\times10^{4}$ & 0.04 &0.02\\ \hline
x&x&&x&& 6.64$\times10^{4}$ & 0.02 &0.04\\ \hline
&x&x&x&x& 6.66$\times10^{4}$ & 0.05 &0.06\\ \hline
&&x&x&x& 6.97$\times10^{4}$ & 0.06 &0.04\\ \hline
x&x&x&x&& 7.40$\times10^{4}$ & 0.07 &0.06\\ \hline
x&&x&x&& 7.58$\times10^{4}$ & 0.07 &0.03\\ \hline
&&x&&& 7.98$\times10^{4}$ & -0.02 &0.00\\ \hline
x&x&&&& 8.18$\times10^{4}$ & 0.04 &0.01\\ \hline
x&x&x&&& 8.80$\times10^{4}$ & 0.07 &0.01\\ \hline
&x&&x&& 8.91$\times10^{4}$ & -0.01 &0.02\\ \hline
&&&x&& 9.12$\times10^{4}$ & 0.00 &0.00\\ \hline
&&x&x&& 9.23$\times10^{4}$ & 0.08 &0.03\\ \hline
&x&x&x&& 9.67$\times10^{4}$ & 0.07 &0.06\\ \hline
&x&&&& 9.80$\times10^{4}$ & 0.00 &0.00\\ \hline
&x&x&&& 1.00$\times10^{5}$ & 0.09 &0.01\\ \hline
x&&&&& 1.07$\times10^{5}$ & 0.00 &0.00\\ \hline
x&&x&&& 1.12$\times10^{5}$ & 0.06 & 0.07
\enddata
\tablecomments{An x indicates that the given parameter is included in the linear regression or logistic regression fit of the stellar obliquity. For each combination of parameters, a Monte Carlo experiment with 1000 iterations is carried out. For each realization, the model is fit to a Monte Carlo realization of 70\% of the data set. The model is then used to make predictions for the remaining 30\% of the data. In the linear regression case, we calculate a modified sum of squared errors that takes into account the measurement uncertainties, sum across the 1000 iterations and then divide by 1000. Models with smaller values are better able to predict the degree of misalignment in the test data. The table is ordered such that the modified sum of squared errors increases toward the bottom.  In the logistic regression case, we calculate the Matthews correlation coefficient, sum across the 1000 iterations, and then divide by 1000. The best-performing model relies on the fundamental stellar parameters included in the modeling scheme. Coefficients closer to 1 indicate that a model can effectively predict the misalignment/alignment classification of systems in the test set. We carry out the logistic regression analysis twice, using the 12.8$^\circ$ cutoff in one experiment and using a 21$^\circ$ cutoff in the second experiment. Using the 12.8$^\circ$ cutoff, the model with the highest MCC includes age, convective envelope mass, and stellar mass. Using the 24$^\circ$ cutoff, the model with the highest MCC includes age, stellar mass, and $\log_{10}(a/R_\ast)$. Overall, in both logistic regression cases, no models are especially effective at predicting the state of misalignment in a system and the relative difference between models is small. We bold the metrics corresponding to the most highly-preferred model in each case.}
\end{deluxetable*}

In the linear regression case, the model that is most effective at predicting the degree of misalignment in a system uses the two fundamental stellar parameters included in our modeling -- stellar age and stellar mass.
 In the logistic regression case, while some models are marginally more effective, all models are not much more effective than if they were assigning classes at random. When using the 12.8$^\circ$ definition of misalignment, the model with the highest MCC includes age, convective envelope mass, and stellar mass. When using the 24$^\circ$ cutoff, the model with the highest MCC includes age, stellar mass, and $\log_{10}(a/R_\ast)$.

For all models considered, we estimate the significance of the coefficients to evaluate the relative importance of each predictor. We proceed using a Monte Carlo experiment similar to the one used in our model selection step. That is, we begin an iteration by sampling the parameters of interest from within their uncertainties. We then fit all data points using the linear or logistic regression model of interest, and retain the $p$-values for each coefficient in the model. After all iterations, we use the median $p$-value for each coefficient as a point estimate for the overall $p$-value for that predictor. 

In Table 4 we show the $p$-values for coefficients in our linear regression models. Lower $p$-values indicate that a change in that independent variable is related to change in the dependent variable. We use .05 as our threshold for 95\% confidence that the coefficient is significant. While $\log_{10}(a/R_\ast)$, $\log_{10}(M_\mathrm{pl}/M_\ast)$ and stellar mass each have a significant coefficient in their respective 1-parameter models, no parameter is significant when fit in conjunction with stellar mass. This suggests that the other parameters only exhibit significant relationships with obliquity due to their collinearity with stellar mass. Overall, this analysis shows that the fraction of the stellar mass in the convective envelope is not the most important parameter in determining misalignment. Instead, the robust significance of $M_\ast$ indicates that it is the most important parameter in predicting the stellar obliquities of hot Jupiter systems.

We show in Table 5 the $p$-values for the logistic regression models tested when using a cutoff of 12.8$^\circ$. As was suggested by the low Matthews correlation coefficients for all models, none of the parameters have a significant coefficient, indicating that none of the models have predictive power. Therefore, this portion of the analysis does not provide insight into what system parameter best predicts the state of misalignment. However, the same is not true when using a cutoff of 24$^\circ$. We show the results for this cutoff value in Table 6. The coefficients for all parameters are significant in their respective 1-parameter models, but only $a/R_\ast$, convective envelope mass, and planet-to-star mass ratio are significant in two parameter models. Overall, $a/R_\ast$ has the most statistically significant coefficient, as well as the smallest p-value in any 2-parameter model including stellar mass, so we argue that it plays the biggest role in determining the state of misalignment in a system.

\begin{deluxetable*}{ccccc}
\tablecaption{Linear Regression Model $p$-values}
\tablenum{4}
\tablewidth{5cm}
\tablehead{\colhead{$t_\ast$} & \colhead{$\log_{10}(a/R_\ast)$} & \colhead{$M_\mathrm{CE}$} & \colhead{$\log_{10}(M_\mathrm{pl}/M_\ast)$} &\colhead{$M_\ast$}}
%% All data must appear between the \startdata and \enddata commands
\startdata
-&-&-&-&$\mathbf{7.35\times10^{-12}}$\\
-&-&-&$\mathbf{6.21\times10^{-11}}$&-\\
-&-&$5.25\times10^{-1}$&-&-\\
-&$\mathbf{8.32\times10^{-10}}$&-&-&-\\
$1.59\times10^{-1}$&-&-&-&-\\ \hline
-&-&-&$8.01\times10^{-1}$&$\mathbf{3.79\times10^{-2}}$\\
-&-&$2.97\times10^{-1}$&-&$\mathbf{6.77\times10^{-12}}$\\
-&$6.09\times10^{-1}$&-&-&$\mathbf{2.21\times10^{-3}}$\\
$3.69\times10^{-1}$&-&-&-&$\mathbf{1.65\times10^{-11}}$\\
-&-&$2.21\times10^{-1}$&$\mathbf{4.82\times10^{-11}}$&-\\
-&$2.72\times10^{-1}$&-&$\mathbf{1.29\times10^{-2}}$&-\\
$2.07\times10^{-1}$&-&-&$\mathbf{9.49\times10^{-11}}$&-\\
-&$\mathbf{7.69\times10^{-10}}$&$3.12\times10^{-1}$&-&-\\
$2.23\times10^{-1}$&-&$6.63\times10^{-1}$&-&-\\
$1.79\times10^{-1}$&$\mathbf{1.08\times10^{-9}}$&-&-&-\\ \hline
-&-&$2.99\times10^{-1}$&$7.93\times10^{-1}$&$5.11\times10^{-2}$\\
-&$4.45\times10^{-1}$&-&$5.42\times10^{-1}$&$5.89\times10^{-2}$\\
$3.54\times10^{-1}$&-&-&$7.70\times10^{-1}$&$6.24\times10^{-2}$\\
-&$6.95\times10^{-1}$&$3.28\times10^{-1}$&-&$\mathbf{2.36\times10^{-3}}$\\
$4.22\times10^{-1}$&-&$3.41\times10^{-1}$&-&$\mathbf{1.39\times10^{-11}}$\\
$4.19\times10^{-1}$&$7.96\times10^{-1}$&-&-&$\mathbf{4.56\times10^{-3}}$\\
-&$2.53\times10^{-1}$&$2.03\times10^{-1}$&$\mathbf{1.01\times10^{-2}}$&-\\
$2.90\times10^{-1}$&$3.94\times10^{-1}$&-&$\mathbf{2.05\times10^{-2}}$&-\\
$2.05\times10^{-1}$&$\mathbf{9.26\times10^{-10}}$&$3.50\times10^{-1}$&-&-\\ \hline
-&$4.11\times10^{-1}$&$2.85\times10^{-1}$&$4.52\times10^{-1}$&$7.97\times10^{-2}$\\
$3.93\times10^{-1}$&-&$3.32\times10^{-1}$&$7.32\times10^{-1}$&$7.40\times10^{-2}$\\
$4.32\times10^{-1}$&$5.46\times10^{-1}$&-&$5.46\times10^{-1}$&$7.45\times10^{-2}$\\
$4.62\times10^{-1}$&$8.09\times10^{-1}$&$3.49\times10^{-1}$&-&$\mathbf{4.35\times10^{-3}}$\\
$3.23\times10^{-1}$&$3.62\times10^{-1}$&$2.36\times10^{-1}$&$\mathbf{1.54\times10^{-2}}$&-\\ \hline
$4.83\times10^{-1}$&$4.75\times10^{-1}$&$3.06\times10^{-1}$&$4.53\times10^{-1}$&$9.80\times10^{-2}$
\enddata
\tablecomments{Each column shows the median $p$-value of the parameter's coefficient across 1000 fits between Monte Carlo realizations of the set of parameters and $|\lambda|$. Those that are significant at the 95\% level are bolded. We use horizontal lines to divide 1-, 2-, 3-, 4-, and 5-parameter models. $M_\ast$ shows the most significant correlation with obliquity and no other parameter is significant when fit in parallel with $M_\ast$. This shows that the significance of the other one-parameter models was due to their collinearity with stellar mass, while stellar mass is the variable that best explains variation in $|\lambda|$.}
\end{deluxetable*}

\begin{deluxetable*}{ccccc}
\tablecaption{Logistic Regression Model $p$-values, 12.8$^\circ$ cutoff}
\tablenum{5}
\tablewidth{10cm}
\tablehead{\colhead{$t_\ast$} & \colhead{$\log_{10}(a/R_\ast)$} & \colhead{$M_\mathrm{CE}$} & \colhead{$\log_{10}(M_\mathrm{pl}/M_\ast)$} &\colhead{$M_\ast$}}
%% All data must appear between the \startdata and \enddata commands
\startdata
% 0.51 & - & - & - & - \\ 
% - & 0.58 & - & - & - \\ 
% - & - & 0.13* & - & - \\
% - & - & - & 0.71 & - \\
% - & - & - & - & 0.67 \\ 
% 0.49 & 0.56 & - & - & - \\
% 0.49 & - & 0.13* & - & - \\
% 0.47 & - & - & 0.65 & -\\
% 0.49 & - & - & - & 0.65\\
% - & 0.46 & 0.12* & - & - \\
% - & 0.17 & - & - & 0.19 \\
% - & - & 0.13* & 0.68 & - \\
% - & - & 0.07* & - & 0.24 \\
% - & - & - & 0.50 & 0.51 \\
% 0.45 & - & 0.07* & - & 0.27\\
-&-&-&-&0.64\\
-&-&-&0.67&-\\
-&-&0.25&-&-\\
-&0.63&-&-&-\\
0.40&-&-&-&-\\ \hline
-&-&-&0.27&0.26\\
-&-&0.17&-&0.35\\
-&0.24&-&-&0.23\\
0.22&-&-&-&0.31\\
-&-&0.15&0.40&-\\
-&0.51&-&0.52&-\\
0.23&-&-&0.37&-\\
-&0.45&0.19&-&-\\
0.59&-&0.33&-&-\\
0.28&0.46&-&-&-\\ \hline
-&-&0.30&0.66&0.62\\
-&0.58&-&0.64&0.31\\
0.43&-&-&0.65&0.55\\
-&0.59&0.31&-&0.46\\
0.50&-&0.37&-&0.27\\
0.43&0.59&-&-&0.39\\
-&0.60&0.20&0.54&-\\
0.24&0.58&-&0.48&-\\
0.53&0.36&0.31&-&-\\ \hline
-&0.62&0.33&0.65&0.58\\
0.55&-&0.35&0.68&0.70\\
0.45&0.60&-&0.64&0.57\\
0.58&0.62&0.39&-&0.49\\
0.50&0.60&0.32&0.49&-\\ \hline
0.56&0.61&0.38&0.64&0.69
\enddata
\tablecomments{Each column shows the median $p$-value of the parameter across 1000 fits between Monte Carlo realizations of the set of parameters and the discrete variable $|\lambda|>12.8^\circ$. We use horizontal lines to divide 1-, 2-, 3-, 4-, and 5-parameter models. None of the parameters have median values of the coefficients significant at the 95\% level. We conclude that none of the models can effectively predict the state of misalignment.}
\end{deluxetable*}

\begin{deluxetable*}{ccccc}
\tablecaption{Logistic Regression Model $p$-values, 24$^\circ$ cutoff}
\tablenum{6}
\tablewidth{10cm}
\tablehead{\colhead{$t_\ast$} & \colhead{$\log_{10}(a/R_\ast)$} & \colhead{$M_\mathrm{CE}$} & \colhead{$\log_{10}(M_\mathrm{pl}/M_\ast)$} &\colhead{$M_\ast$}}
%% All data must appear between the \startdata and \enddata commands
\startdata
-&-&-&-&$\mathbf{2.01\times10^{-2}}$\\
-&-&-&$\mathbf{6.16\times10^{-3}}$&-\\
-&-&$\mathbf{8.68\times10^{-3}}$&-&-\\
-&$\mathbf{3.35\times10^{-3}}$&-&-&-\\
$\mathbf{1.26\times10^{-2}}$&-&-&-&-\\ \hline
-&-&-&$\mathbf{4.38\times10^{-2}}$&$1.67\times10^{-1}$\\
-&-&$\mathbf{3.83\times10^{-2}}$&-&$3.05\times10^{-1}$\\
-&$\mathbf{2.19\times10^{-2}}$&-&-&$1.57\times10^{-1}$\\
$8.24\times10^{-2}$&-&-&-&$5.43\times10^{-1}$\\
-&-&$1.02\times10^{-1}$&$2.40\times10^{-1}$&-\\
-&$1.73\times10^{-1}$&-&$4.22\times10^{-1}$&-\\
$2.18\times10^{-1}$&-&-&$4.10\times10^{-1}$&-\\
-&$1.66\times10^{-1}$&$1.53\times10^{-1}$&-&-\\
$2.94\times10^{-1}$&-&$1.89\times10^{-1}$&-&-\\
$3.12\times10^{-1}$&$2.53\times10^{-1}$&-&-&-\\ \hline
-&-&$2.35\times10^{-1}$&$4.13\times10^{-1}$&$5.53\times10^{-1}$\\
-&$2.74\times10^{-1}$&-&$6.30\times10^{-1}$&$2.23\times10^{-1}$\\
$4.77\times10^{-1}$&-&-&$2.53\times10^{-1}$&$3.48\times10^{-1}$\\
-&$1.94\times10^{-1}$&$3.06\times10^{-1}$&-&$3.72\times10^{-1}$\\
$5.02\times10^{-1}$&-&$1.91\times10^{-1}$&-&$5.57\times10^{-1}$\\
$4.95\times10^{-1}$&$1.31\times10^{-1}$&-&-&$2.42\times10^{-1}$\\
-&$3.04\times10^{-1}$&$1.58\times10^{-1}$&$4.62\times10^{-1}$&-\\
$2.83\times10^{-1}$&$2.44\times10^{-1}$&-&$3.95\times10^{-1}$&-\\
$6.08\times10^{-1}$&$3.14\times10^{-1}$&$2.44\times10^{-1}$&-&-\\ \hline
-&$3.00\times10^{-1}$&$3.24\times10^{-1}$&$6.59\times10^{-1}$&$5.57\times10^{-1}$\\
$5.81\times10^{-1}$&-&$3.02\times10^{-1}$&$5.08\times10^{-1}$&$6.03\times10^{-1}$\\
$5.12\times10^{-1}$&$2.71\times10^{-1}$&-&$6.66\times10^{-1}$&$4.35\times10^{-1}$\\
$6.05\times10^{-1}$&$2.47\times10^{-1}$&$3.78\times10^{-1}$&-&$3.86\times10^{-1}$\\
$5.58\times10^{-1}$&$3.22\times10^{-1}$&$2.88\times10^{-1}$&$4.49\times10^{-1}$&-\\ \hline
$5.56\times10^{-1}$&$2.95\times10^{-1}$&$3.67\times10^{-1}$&$6.50\times10^{-1}$&$6.27\times10^{-1}$
\enddata
\tablecomments{Each column shows the median $p$-value of the parameter across 1000 fits between Monte Carlo realizations of the set of parameters and the discrete variable $|\lambda|>12.8^\circ$. We use horizontal lines to divide 1-, 2-, 3-, 4-, and 5-parameter models. The coefficients for all parameters are significant in their respective 1-parameter models, but only $\log_{10}(a/R_\ast)$, convective envelope mass, and $\log_{10}(M_\mathrm{pl}/M_\ast)$ are significant in two parameter models. Overall,  $\log_{10}(a/R_\ast)$ has the most statistically significant coefficient, as well as the smallest p-value in any 2-parameter model including stellar mass.}
\end{deluxetable*}
\section{Discussion}
We have shown that the Galactic velocity dispersion of hosts of misaligned hot Jupiters is typically larger than the Galactic velocity dispersion of hosts of aligned hot Jupiters. This observation suggests that misaligned systems are older than aligned systems, in contrast to what would be expected if a single formation mechanism produces hot Jupiters with a wide range of stellar obliquities which are then damped by tides. Additionally, we have found that stellar mass plays the most significant role in predicting the degree of misalignment in a system, and that depending on the delineation between aligned and misaligned systems, $a/R_\ast$ may play a significant role in predicting whether a system will be misaligned. 

Now, we examine the assumptions made in our analysis and how they affect the interpretation of our results. In our analysis, we use the sky-projected stellar obliquity, $\lambda$, as a proxy for the true stellar obliquity, $\psi$, in most systems. Figure 4 of \citet{Fabrycky2009} shows the probability density for $\psi$ conditional on an observed $\lambda$. In the case of a small value for $\lambda$, there is a broad tail toward larger values for $\psi$, and therefore the sky-projected value is a lower limit on the true obliquity. In the case of a large $\lambda$, the tail is not as broad, and so the sky-projected stellar obliquity is likely close to the true value. Overall, this means that it is most likely that our aligned samples are more contaminated with misaligned systems than the reverse. This does not affect the general results of our analysis, but likely has the effect of lowering the significance of our results from what we would have recovered had we been working with the true obliquity in all systems.

An assumption inherent in our analysis is that the tidal realignment process will result in prograde systems, in which $\lambda=0^\circ$. In particular, we search for evidence of tidal realignment by comparing the Galactic velocity dispersions of aligned systems and misaligned systems when defining alignment relative to $\lambda =  12.8^\circ$. We define the aligned and misaligned classes in the logistic regression analysis in the same way. \citet{Rogers2013b} find that while initially prograde hot Jupiters evolve toward $0^\circ$, damping of inertial waves causes initially retrograde hot Jupiters align toward $90^\circ$ or $180^\circ$. It may also be that the process of tidal realignment can stall at 90$^\circ$. Theoretical work suggests that this would be the case when the dominant mode of tidal dissipation is the dissipation of inertial waves in the convective envelope of Sun-like stars \citep[e.g.,][]{Lai2012}. However, the dissipation of inertial waves requires that the tidal forcing frequency in the rotating frame is less than twice the stellar spin frequency, and this condition is not satisfied for most of the systems in our sample in their present day configurations. Moreover, \citet{Xue2014} and \citet{Li2016} find that when equilibrium tides and magnetic braking are accounted for, the systems which stall at $90^\circ$ will eventually align to prograde orbits. Hence, our assumption may be sound. On the other hand, recent observational studies have identified a preference for perpendicular orbits among misaligned systems. Specifically, \citet{Albrecht2021} examine a subset of systems for which the true obliquity can be constrained and find an overabundance of planets with near-polar orbits relative to what one would expect if stellar obliquities were randomly oriented. Therefore, it is ambiguous whether or not prograde-aligned systems are the generic result of tidal evolution. Despite this, we do not think there is a better assumption to make when searching for evidence of realignment given the relatively large uncertainties on $\lambda$, the fact that we are restricted to the sky-projected value in most cases, and the uncertainties about tidal theory.

\subsection{Galactic Velocity Dispersion Analysis}
We have shown that the Galactic velocity dispersion of the hosts of misaligned hot Jupiters is typically larger than the Galactic velocity dispersion of the hosts of aligned hot Jupiters. This offset is highly significant for the subsample of stars with smaller-than-median fractional convective envelope masses, and the offset is present but statistically insignificant for the subsample of stars with larger-than-median fractional convective envelope masses. This is counter to what would be expected if hot Jupiters are formed by a single process that generates them with a wide range of obliquities and misaligned hot Jupiters are realigning the photospheres of their host stars via tidal interactions. This observation appears to imply that systems with misaligned hot Jupiters are older than systems with aligned hot Jupiters, but this signal may be a product of other effects which we rule out below. 

\begin{figure*}[h]
    \centering
    \plotone{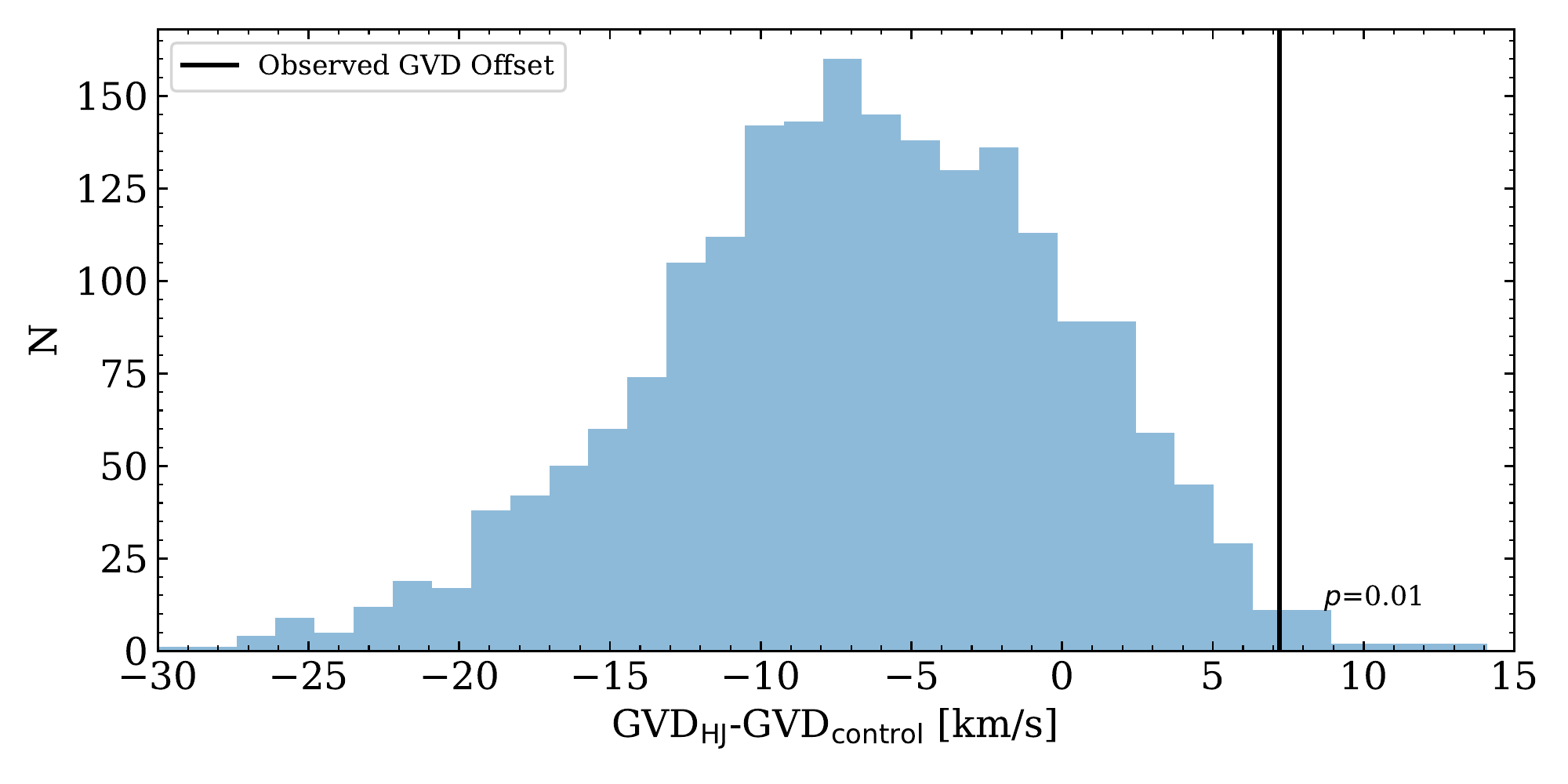}
    \caption{Difference in Galactic velocity dispersion between samples of hot Jupiter hosts and matched field stars. The sample size is matched to that of the number of aligned systems hosted by stars with smaller-than-median fractional convective envelope masses. The black line indicates the difference between the median Galactic velocity dispersions of the misaligned and aligned system populations hosted by stars with smaller-than-median fractional convective envelope masses. Less than 1\% of the hot Jupiter host star samples and matched control samples exhibit a Galactic velocity dispersion offset larger than that of the misaligned and aligned systems hosted by stars with smaller-than-median fractional convective envelope masses. Therefore, we argue that it is likely that the misaligned population genuinely has a larger Galactic velocity dispersion than the aligned population hosted by stars with smaller-than-median fractional convective envelope masses, implying a larger relative age for the misaligned population.}
    \label{fig:Figure 5}
\end{figure*}

It is possible that the offset we observe is a consequence of the small samples we are working with. We carry out a test to determine if the Galactic velocity dispersion offsets we observed are likely to arise due to the small size of the subsamples being compared. In other words, if we assume that at the population level the aligned systems are older than the misaligned systems, how often would random subsamples exhibit an offset between their Galactic velocity dispersions at least as large as the offsets observed in the top and bottom panels of Figure~\ref{fig:Figure 4}? Obviously, we do not have knowledge of the overall population that our observed sample is selected from. In place of this, we substitute two samples of stars known to exhibit different relative ages. \citet{Hamer2019} showed that hot Jupiter host stars typically have a smaller velocity dispersion than matched samples of stars not hosting hot Jupiters. In the top panel of Figure~\ref{fig:Figure 4}, the misaligned systems have a median Galactic velocity dispersion larger than that of the aligned systems by 7.5 km/s. The smaller sample of the two typically has 20 members. Over 5000 iterations, we select 20 hot Jupiter host stars and a matched sample of 20 field stars without hot Jupiters from the hot Jupiter host star and field star samples used in \citet{Hamer2019} and using the same matching procedure as in that analysis. We record the difference between the Galactic velocity dispersion of the hot Jupiter host star sample and that of the matched field star sample. We then determine how often the difference is larger than the offset in the top panel of Figure~\ref{fig:Figure 4}. The result of this experiment is shown in Figure~\ref{fig:Figure 5}. The experiment shows that for stars with smaller-than-median fractional convective envelope masses, there is less than a 1\% chance that the observed offset does not reflect an offset present at the population level. Overall, it is unlikely that the Galactic velocity dispersion offset we observe is a consequence of the small sample sizes.

We note that the above experiments rely on the assumption that tidal realignment takes place on a similar timescale to orbital decay. This is because the cause for the difference in relative age between the hot Jupiter host star and field star samples is the orbital decay of hot Jupiters during their host stars' main sequence lifetimes. If realignment takes place on a shorter timescale than orbital decay, then any age differences between misaligned and aligned systems would be smaller than those corresponding to age differences related to orbital decay. Our method would not be able to recover these age differences even with a sample large enough to recover age differences induced by orbital decay. 

\begin{figure*}[h]
    \centering
    \plotone{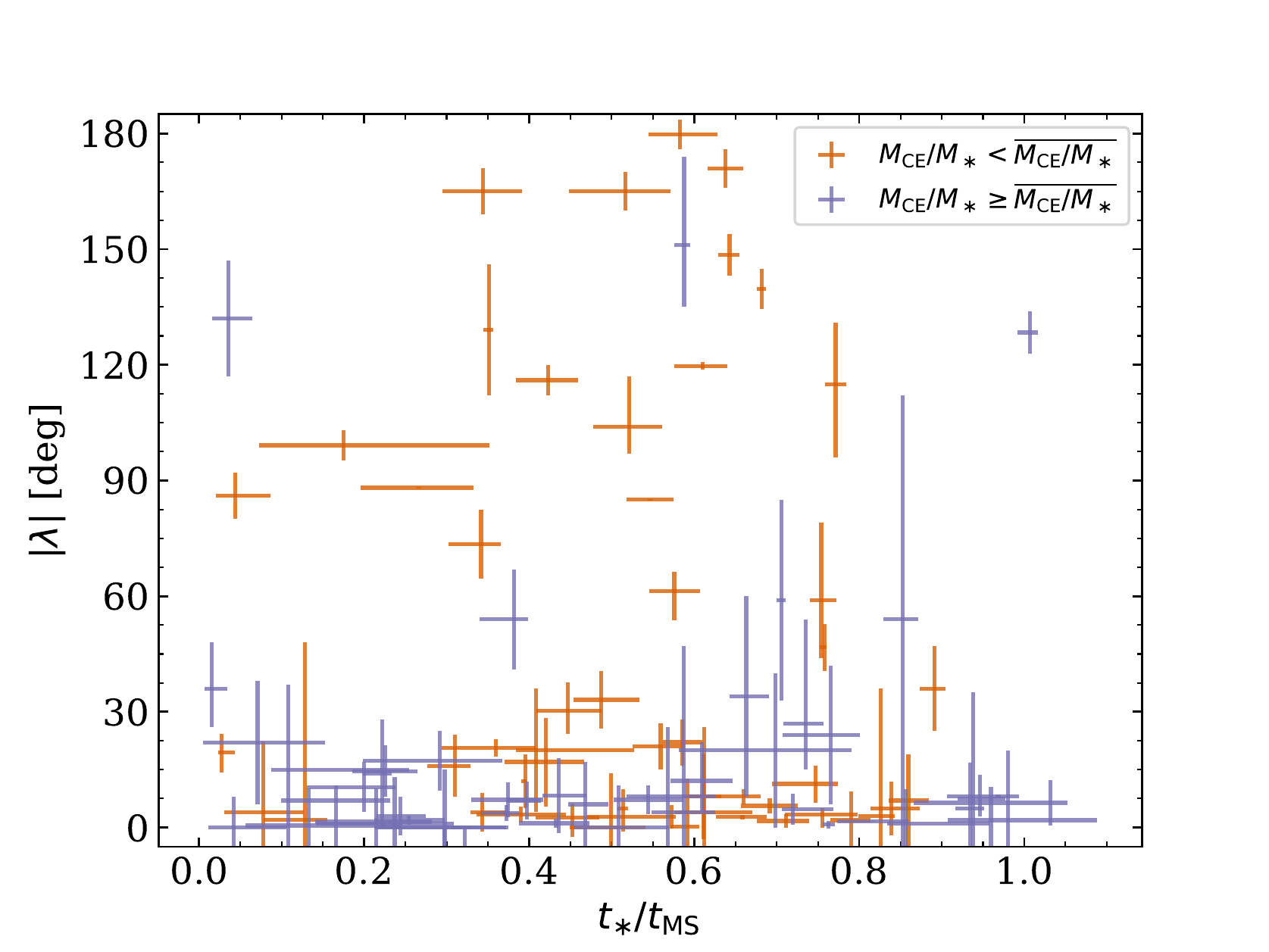}
    \caption{Stellar obliquities of hot Jupiter host stars plotted against the host stars' isochronal ages scaled to their main sequence lifetimes. The main sequence lifetimes of the stars in the sample are determined using terminal age main sequence of the MIST evolutionary tracks for a star of that mass and metallicity. Stars with larger-than-median fractional convective envelope masses are shown as blue points, and stars with smaller-than-median fractional convective envelope masses are shown as orange points. Misalignments are present even for stars which have lived out a significant portion of their main sequence lifetimes. Hence, it is unlikely that tidal realignment is a significantly faster process than tidal orbital decay.}
    \label{fig:Figure 6}
\end{figure*}

We show that tidal realignment cannot be occurring significantly faster than orbital decay, as misalignments persist for a significant fraction of the main sequence lifetimes of the host stars in our sample. We determine the main sequence lifetime of our host stars using the terminal age main sequence of the MIST evolutionary tracks. Then, we scale the isochronal ages of our host stars by these main sequence lifetime estimates. We show the result of this calculation in Figure~\ref{fig:Figure 6}.  Misalignments are present even within systems where a significant portion of the host star's main sequence lifetime has elapsed. One may be concerned that the systems with significant misalignments that have persisted throughout a large fraction of the host stars' main sequence lifetimes simply have a configuration which causes tidal realignment to be negligible. But we compare the $a/R_\ast$ and planet-to-star mass ratio distributions of the overall sample to the subsamples with larger than average misalignments and larger than average $t_\ast/t_\mathrm{MS}$ and find that this subsample has similar distributions of these parameters which set the alignment timescale. Hence, we argue that tidal realignment cannot occur on a timescale much more quickly than that of tidal orbital decay. Moreover, the Galactic velocity dispersion offset we are considering in our sample size argument is amongst the aligned and misaligned populations hosted by stars with smaller-than-median fractional convective envelope masses. It is not expected that tidal realignment can be a fast process in systems hosted by these stars, if it even operates at all, so it is safe to assume that the realignment timescale is not much shorter than that of orbital decay. Moreover, even if realignment occurred rapidly, then we would expect the aligned systems and misaligned systems to have consistent Galactic velocity dispersions rather than exhibiting the offset we observe.

\begin{figure*}
    \centering
    \plotone{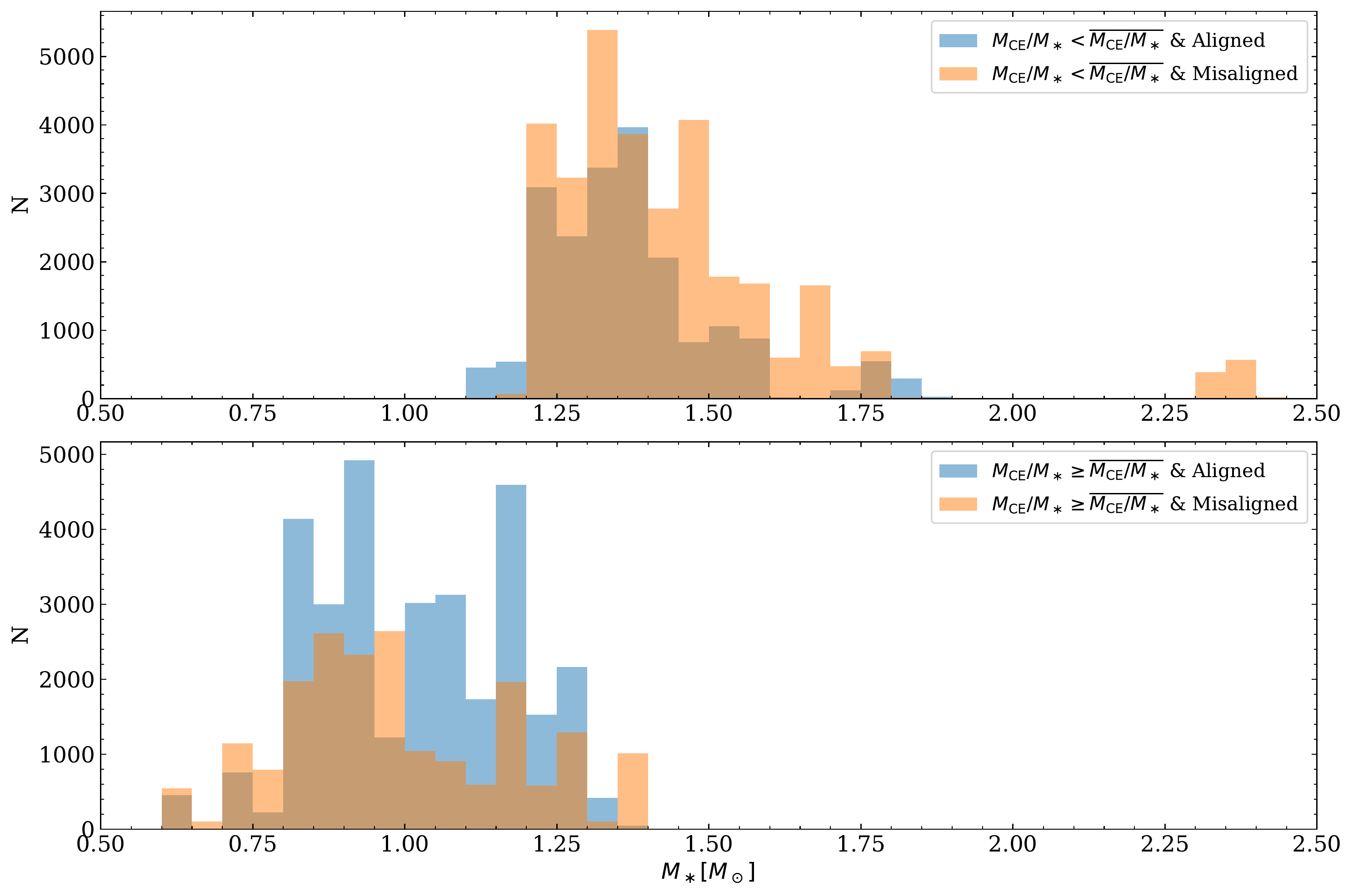}
    \caption{Stellar mass distributions of the host stars in misaligned systems (orange) and the host stars in aligned systems (blue) across 1000 Monte Carlo realizations of the sample. As the mass distribution is consistent between the hosts of misaligned systems and hosts of aligned systems, the difference in the Galactic velocity dispersions of the two samples is not related to stellar mass. Top: Systems in which the host star has smaller-than-median fractional convective envelope mass. Bottom: Systems in which the host star has larger-than-median fractional convective envelope mass.}
    \label{fig:Figure 7}
\end{figure*}

\begin{figure*}
    \centering
    \plotone{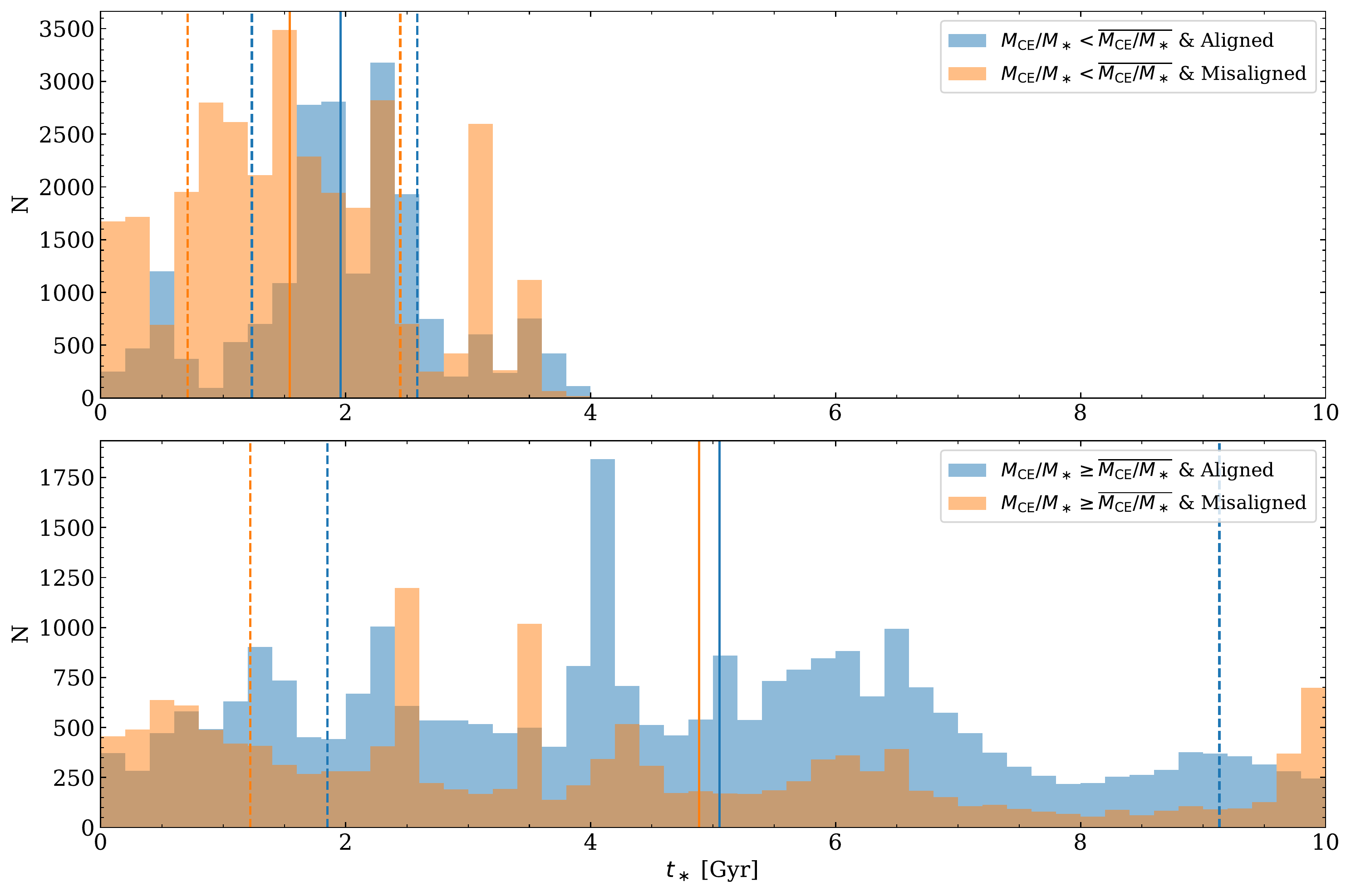}
    \caption{Stellar age distributions of the host stars in misaligned systems (orange) and the host stars in aligned systems (blue) across 1000 Monte Carlo realizations of the sample. Solid lines indicate the median stellar age of each sample. Dotted lines indicate the 16th and 84th percentiles of the stellar age distributions. The misaligned and aligned populations have consistent age distributions. The stellar age inferences from isochrones are not precise enough to reproduce the relative age differences implied by the Galactic velocity dispersion analysis. Top: Systems in which the host star has smaller-than-median fractional convective envelope mass. Bottom: Systems in which the host star has larger-than-median fractional convective envelope mass.}
    \label{fig:Figure 8}
\end{figure*}
It may be that the apparent difference in the relative ages of the subsamples is a consequence of different stellar mass distributions within each subsample. We use a Monte Carlo simulation to rule out this possibility. Simultaneously, we investigate whether or not the isochronal stellar ages corroborate the implications of the Galactic velocity dispersion offsets. We perform a Monte Carlo simulation similar to the one used in the calculation for Figure~\ref{fig:Figure 4}. We sample the stellar obliquities, stellar masses, and stellar ages from within their uncertainty distributions and divide the sample into misaligned and aligned systems across 1000 iterations. We further divide the samples based on whether the fractional convective envelope mass is greater than or less than the median value within the sample. Finally, we compare the distribution of stellar masses and stellar ages for the aligned and misaligned systems to determine if the Galactic velocity dispersion difference is a product of different stellar mass distributions. We show the result of the stellar mass comparison in Figure~\ref{fig:Figure 7}. The stellar mass distribution is consistent for the hosts of misaligned systems and the hosts of aligned systems for both the systems in which the host star has a fractional convective envelope mass smaller than or larger than the median value within the sample. In Figure~\ref{fig:Figure 8}, we show the distribution of stellar ages within each sample. The age distributions of the misaligned and aligned systems are consistent for both of the subsamples. This is due to the imprecision of isochronal age estimates for main sequence stars, and shows the advantage of inferring the relative age of stellar populations with Galactic velocity dispersion comparisons.

%Additionally, the Galactic velocity dispersion offsets we measured could be produced if there is an observational bias toward being able to measure misalignments around older stars. 
We must also consider the possibility that the Galactic velocity dispersion offset between the misaligned and aligned systems is a product of an observational bias. Specifically, this offset could be produced if we are more likely to observe misaligned hot Jupiters orbiting stars that are older or if we are more likely to observe aligned hot Jupiters orbiting stars that are younger. However, we argue that this is unlikely. The RM effect is stronger for faster-rotating stars, meaning that we are more likely able to constrain spin-orbit angles for younger stars, not older ones. Additionally, the amplitude of the RM effect is stronger for smaller stellar radii. As stellar radii grow over time due to stellar evolution, this also means that the bias inherent in being able to constrain the spin-orbit angle is toward younger systems rather than older ones. While there is a preference toward Doppler discovery of hot Jupiters orbiting more slowly rotating stars, this bias is present in both the misaligned and aligned systems, and therefore should not affect this comparison. Additionally, as described in \citet{Winn2010}, there is a bias toward the confirmation of transit-discovered misaligned systems as misaligned systems will have less Doppler broadening than aligned ones. Overall, it is unlikely that the Galactic velocity dispersion offset is the product of an observational bias. 

\begin{figure*}
    \centering
    \plotone{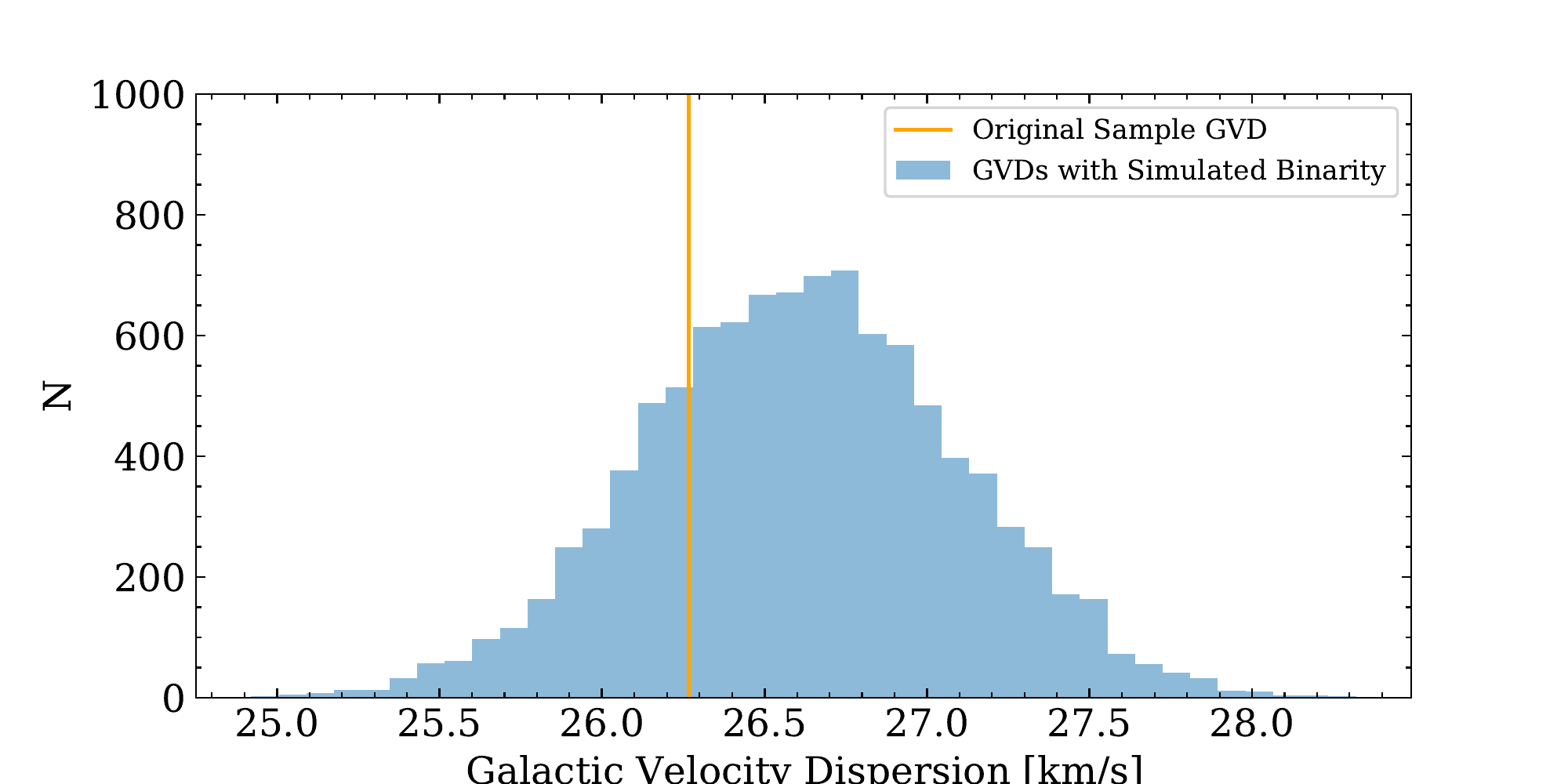}
    \caption{Distribution of Galactic velocity dispersions for a sample of aligned hot Jupiter host stars with smaller-than-median fractional convective envelope masses after simulating the effect of binarity across Monte Carlo realizations. The binary companions are placed at 40 AU to reveal the upper limit of the effect of binarity on the Galactic velocity dispersion of the sample, and because a companion at a smaller separation can suppress the formation of giant planets \citep[e.g.,][]{Kraus2012}. The mass of the companion is chosen such that the companion is 4 magnitudes fainter than the median primary star in this sample in accordance with the fact that these stars have not been detected as double-lined spectroscopic binaries. Across 10000 iterations, the median increase in the Galactic velocity dispersion due to binarity is 0.34 km/s and the maximum increase is 2.1 km/s. As the Galactic velocity dispersions of the hosts of misaligned hot Jupiters is larger than that of the hosts of aligned hot Jupiters by $\sim7.2$ km/s for stars without convective envelopes and $\sim4.6$ km/s for stars with convective envelopes, we argue that unresolved binarity cannot explain the Galactic velocity dispersion offsets we observed.}
    \label{fig:Figure 9}
\end{figure*}

%It could also be that the hosts of misaligned systems have a higher binary companion fraction, inflating their Galactic velocity dispersion.
It is possible that the Galactic velocity dispersion offset between the hosts of misaligned systems and hosts of aligned systems is a product of a higher companion fraction for the hosts of misaligned systems. In order to determine if the offset can be explained by binarity, we carry out the following experiment. As we want to determine how much binarity can inflate the Galactic velocity dispersion of a population of stars, we simulate the effect of binarity on a subsample of aligned hot Jupiter host stars with smaller-than-median fractional convective envelope masses, supposing that their lower typical Galactic velocity dispersions are the result of those host stars more often being single. As these stars have not been detected as double-lined spectroscopic binaries, any possible companions must be 3-5 magnitudes fainter. We use MIST isochrones to determine the stellar mass at which a companion star would be four magnitudes fainter than the primary star at the median stellar mass, age, and metallicity in the subsample ($M_\ast$ = 1.33 $M_\odot$, $t_\ast$ = $10^{9.26}$ yr, [Fe/H]=0.10). Such a companion star would have $M_2=0.68\ M_\ast$. In order to determine an upper limit on the inflation of the Galactic velocity dispersion by binarity, we place this companion on a circular orbit at 40 AU from each host star and calculate the semiamplitude of the reflex motion it imparts on the primary star. We place the companion at 40 AU because observations have shown that binary companions at a separation of less than 40 AU cause protoplanetary discs to disperse in $\lesssim$1 Myr, making giant planet formation unlikely in any binary system with a smaller separation \citep[e.g.,][]{Kraus2012}. 
For each of these binaries, we generate an inclination betweeen $(0^\circ,90^\circ)$ according to an isotropic distribution of orbital alignments and sample the phase of the orbit from a uniform distribution between $(0^\circ,360^\circ)$ across 10000 iterations. At each iteration, we perturb the radial velocity and proper motions of each star according to the induced reflex motion, inclination, and orbital phase and then recalculate the Galactic velocity dispersion. The result of this Monte Carlo experiment is shown in Figure~\ref{fig:Figure 9}. The median increase in Galactic velocity dispersion is only $\sim0.34$ km/s, and the maximum increase in Galactic velocity dispersion across the 10000 iterations is just $\sim2.1$ km/s. Hence, we argue that binarity cannot account for the magnitude of the Galactic velocity dispersion offsets we observe between our misaligned and aligned host star samples, which are $\sim7.2$ km/s for the stars with smaller-than-median convective envelope masses and $\sim4.6$ km/s for the sample of stars with larger-than-median fractional convective envelope masses. If we instead assume the companions are of equal mass with the primary to determine the upper limit of the effect of binarity, an unreasonable assumption given that the putative companion stars have not been detected, then the median increase in the Galactic velocity dispersion is $\sim1.9$ km/s and the maximum increase is $\sim5.8$ km/s.  Our finding that the Galactic velocity dispersion offsets cannot be explained by a higher binary fraction agrees well with the finding that hosts of misaligned hot Jupiters have a companion fraction consistent with that of the overall population of hot Jupiter hosts \citep{Ngo2015}.

We have explored every explanation for the observed Galactic velocity dispersion offset other than the possibility that it reflects a genuine age difference between the aligned and misaligned populations. The offset suggests that misaligned systems are relatively older than aligned systems, which is inconsistent with the hypothesis that hot Jupiters are generated with a wide range of obliquities by a single process and that aligned systems are produced by tidal realignment. 

\subsection{The Broader Context of Stellar Obliquities}
Our Galactic velocity dispersion analysis showed that the relative ages of misaligned hot Jupiters and aligned hot Jupiters are inconsistent with the hypothesis that hot Jupiters are formed with a wide range of stellar obliquities and that aligned systems are produced when hot Jupiters have realigned the photospheres of their host stars. Therefore, we consider if alternative explanations for the apparent pattern in hot Jupiter stellar obliquities can explain the observation that misaligned systems are older than aligned systems.
%Given the significant correlation between stellar mass and the degree of misalignment, we argue that the pattern must be related to some other process mediated by the change in stellar structure at $M_\ast\sim1.2M_\odot$. 
 
The pattern can be produced by different angular momentum transport processes in low- and high-mass stars. Specifically, \citet{Rogers2012, Rogers2013} showed that internal gravity waves may significantly modulate the rotation of hot star photospheres over astrophysically short timescales. But this process is expected to significantly change the rotation rate of the photosphere over just hundreds or thousands of spin periods -- much shorter than the main sequence lifetime of a star and likely too small of an age difference to detect with our method of comparing Galactic velocity dispersions. 

Alternatively, the pattern could be produced if protoplanetary disks are misaligned from the spin of their host stars, and then are preferentially realigned only in systems hosted by cool stars. 
Misalignments between stellar and protoplanetary disk angular momenta can be generated in a variety of physical processes, resulting in planetary systems with orbital angular momenta misaligned with their host stars' rotational angular momenta. Molecular clouds are turbulent, so the orientation of high angular momentum material accreting at relatively late times onto the outer edge of the protoplanetary disk may not share the same angular momentum orientation as material already part of the star that accreted at early times. A protoplanetary disk's angular momentum can therefore become misaligned with the rotational axis of the star growing at its center \citep[e.g.,][]{Bate2010,Fielding2015}. Additionally, an initially aligned disk could be torqued out of alignment by a stellar mass \citep[e.g.,][]{Batygin2012, Lai2014, Spalding2014} or planetary mass \citep[e.g.,][]{Matsakos2017} companion. While the disk could become misaligned around any type of star, \citet{Spalding2015} showed that only lower-mass stars have magnetic fields strong enough to realign the angular momenta of the stellar spin and the protoplanetary disk during the disc's lifetime. This scenario could also account for the fact that Kepler planets exhibit a similar pattern as the one we observe in the stellar obliquities of hot Jupiters \citep[e.g.,][]{Mazeh2015, Louden2021}. Lower-mass planets on wider orbits should experience negligible tidal interactions with their hosts, so it is clear that tides are not the cause of the pattern for these smaller, longer-period planets. But if misalignments are generated at the protoplanetary disk phase, then the pattern in the obliquities would be in place by the time the protoplanetary disk dissipates. So in this scenario, one would expect that aligned systems and misaligned systems would have consistent age distributions rather than the offset we observe.

If the processes that produce the pattern in the stellar obliquities of hot Jupiters and small planets would not result in the age difference we inferred, then the only explanation for this age difference is that some misaligned hot Jupiters form later than aligned hot Jupiters. Either these planets arrived at their present locations early on, and then they were driven to misalignment at late times, or these planets spent a significant amount of time at larger distances from their host stars before migrating to their present close-in, misaligned orbits at late times. Our results are consistent with the findings of \citet{Spalding2022}, who show that if misaligned hot Jupiters orbiting hot stars must form late, as if they had formed early then they would have been tidally realigned when their host stars had convective envelopes during the pre-main sequence.

The late-forming misaligned hot Jupiters must be formed by a process that generates misalignments and takes place over relatively long timescales. This likely rules out their formation via disk migration in a misaligned protoplanetary disk \citep[e.g.,][]{Lin1996, Spalding2014}, or by planet-planet scattering occurring soon after the dispersal of the protoplanetary disk \citep[e.g.,][]{Chatterjee2008}. Instead, these planets are most likely produced by secular processes that take place over many orbital periods. For example, secular interactions between planets following perturbations by a passing star may lead to high eccentricity migration followed by tidal circularization \citep[e.g.,][]{Zakamska2004, Hamers2017, Wang2021, Rodet2021}. Additionally, an inclined outer planet \citep[e.g.,][]{Naoz2011, Teyssandier2013} or stellar companion \citep[e.g.,][]{Wu2003,Fabrycky2007,Naoz2012, Petrovich2015, Anderson2016, Vick2019} can raise the eccentricity of a proto-hot Jupiter through von Zeipel-Lidov-Kozai oscillations \citep{vonZeipel1910, Lidov1962,Kozai1962}, triggering high eccentricity migration followed by tidal circularization. Even in a system with moderate eccentricities and inclinations, secular chaos can lead to the formation of misaligned hot Jupiters over long timescales \citep[e.g.,][]{Wu2011, Teyssandier2019}.

In contrast, the early-forming aligned hot Jupiters and any early-forming misaligned hot Jupiters must have been formed through relatively fast processes. As protoplanetary disks may become misaligned from the rotational angular momentum of their central star, early formation of both misaligned and aligned hot Jupiters can be explained by in situ formation \citep[e.g.,][]{Boley2016, Batygin2016, Bailey2018} or by disk migration \citep[e.g.,][]{Lin1996, Ward1997, Ida2004, Kley2012, Heller2019}. Alternatively, aligned hot Jupiters may be able to form relatively quickly through coplanar high-eccentricity migration \citep{Petrovich2015b}. The discovery of an aligned hot Jupiter-size planet orbiting a star just 17 Myr old bolsters the case that hot Jupiters are able to form rapidly, especially those which are aligned \citep{Rizzuto2020, Heitzmann2021}.

While we are arguing that there are multiple formation channels for hot Jupiters, we are not arguing that different formation channels within systems hosted by lower- and higher-mass stars is the cause of the apparent relationship between the sky-projected stellar obliquities of hot Jupiters and the mass/effective temperature of their host stars. Our observation that the population of hosts of misaligned hot Jupiters are relatively older than the hosts of aligned hot Jupiters only requires that some of misaligned hot Jupiters form late, not all. As both aligned and misaligned hot Jupiters can form early through the processes described above, our argument that there are multiple formation channels does not imply that these formation channels can explain the sharp transition in the distribution of stellar obliquities at $\sim1.2\ M_\odot$. 

Our conclusion that the relative ages of misaligned and aligned hot Jupiters disfavored the tidal realignment had supposed that the misalignments in hot Jupiter systems were all primordial. But if some misaligned hot Jupiters are forming late, then the signature of tidal realignment may be masked. This would explain why other studies have found evidence in support of the tidal realignment hypothesis, while our observation initially appears to contradict it \citep[e.g.,][]{Triaud2011, Hebrard2011b, Albrecht2012, Dawson2014, Valsecchi2016, Rice2022}. The reduced significance of the Galactic velocity dispersion offset within the subsample of systems hosted by stars with convective envelopes may itself be a signal of tidal alignment. If some of the late-forming misaligned hot Jupiters go on to realign the photospheres of their host stars with convective envelopes, then this would reduce the age difference between the aligned and misaligned populations. In other words, aligned hot Jupiters hosted by stars with convective envelopes can be produced either very early on or much later through high eccentricity migration followed by tidal realignment. 

Moreover, this could also explain why our linear regression did not identify $a/R_\ast$ as the most important parameter in determining the degree of misalignment despite the strong dependence of the tidal realignment timescale on it. In other words, a relationship between $a/R_\ast$ and $|\lambda|$ introduced by tidal interactions may be washed out by the presence of systems that have been placed at their present locations relatively recently. This ''washing-out" of the signal of tidal realignment would also explain why \citet{Safsten2020} had found that stellar obliquities are best predicted by stellar effective temperature rather than age, as we had. On the other hand, $a/R_\ast$ had the most significant coefficient in the logistic regression model when using a threshold of $24^\circ$ for misalignment, indicating that there may indeed be a relationship between $a/R_\ast$ and $|\lambda|$ that has been weakened out by late-arriving misaligned planets that have been acted on by tides for a shorter amount of time.  

It is also possible that the relationships between system parameters and obliquity we predicted would result from tides were not present due to the complexities of tidal theory. For example, it may be that the relationship between $a/R_\ast$ and obliquity is confounded by the period dependence of $Q_\ast'$ such as the one suggested by \citet{Penev2018}. And while we had predicted that the signature of tidal realignment mediated by the presence of a convective envelope would manifest as a significant correlation between convective envelope mass and obliquity, the picture may be more complicated than that. Specifically, the tidal dissipation via inertial waves may scale with the surface area of the radiative convective boundary \citep[e.g.,][]{Goodman2009}, such that dissipation increases with decreasing convective envelope mass. Overall, the results of our Galactic velocity dispersion analysis, our model selection scheme, and our regression analyses can still be consistent with the scenario of tidal realignment.

\section{Conclusion}
It has been shown that hot Jupiter systems hosted by hot stars exhibit a wider range of spin-orbit angles than those hosted by cool stars. We homogeneously derived stellar parameters and convective envelope masses for host stars of hot Jupiters with reliable inferences of their stellar obliquities. 
We searched for evidence that the pattern in hot Jupiters' stellar obliquities is caused by tidal interactions sculpting a primordial distribution of misalignments, and that these tidal interactions are only efficient in stars with convective envelopes. If hot Jupiters form through a single process that generates a wide range of obliquities and aligned hot Jupiters are the product of tidal realignment, then aligned hot Jupiters should be relatively older than misaligned hot Jupiters. Moreover, if tidal interactions mediated by the convective envelopes of the host stars are producing the observed pattern in the obliquities, then convective envelope mass and the system parameters that set the strength of tidal interactions should show significant correlations with the degree of misalignment in a system. We used the Galactic velocity dispersion as a proxy for relative age and found evidence that for stars with smaller-than-median fractional convective envelope masses, hosts of misaligned hot Jupiters are significantly older than hosts of aligned hot Jupiters. This observation suggests that misaligned hot Jupiters form late relative to aligned hot Jupiters. In contrast, for stars with larger-than-median fractional convective envelope masses, hosts of misaligned hot Jupiters are only marginally older than hosts of aligned hot Jupiters. We interpret this decrease in the significance of the age offset as further evidence that for stars with convective envelopes, hot Jupiters are tidally realigning the photospheres of their host stars. Additionally, we found that stellar mass or scaled semimajor axis play the most significant role in predicting the degree of misalignment in a system, rather than convective envelope mass. Together, these results suggest that misaligned hot Jupiters form late relative to aligned hot Jupiters, and that in systems hosted by stars with convective envelopes tidal realignment may subsequently generate aligned systems. 
%Any possible signal related to tidal realignment may be masked by the fact that the stellar obliquities of hot Jupiters are not all primordial. 
%This is in contrast with models which predict that the convective envelope mass explains the observed pattern in the spin-orbit misalignments of hot Jupiters. 
\section*{Acknowledgements}
%\begin{acknowledgments*}
We thank Josh Winn, Natalie Allen, Cicero Lu, and Zafar Rustamkulov for helpful comments on this paper. This material is based upon work supported by
the National Science Foundation under grant number 2009415.

This work has made use of data from the European Space Agency (ESA) mission {\it Gaia} (\url{https://www.cosmos.esa.int/gaia}), processed by the {\it Gaia} Data Processing and Analysis Consortium (DPAC, \url{https://www.cosmos.esa.int/web/gaia/dpac/consortium}). Funding for the DPAC has been provided by national institutions, in particular the institutions participating in the {\it Gaia} Multilateral Agreement. 

This publication makes use of data products from the Wide-field Infrared Survey Explorer, which is a joint project of the University of California, Los Angeles, and the Jet Propulsion Laboratory/California Institute of Technology, funded by the National Aeronautics and Space Administration. 

This research is based on observations made with the Galaxy Evolution Explorer, obtained from the MAST data archive at the Space Telescope Science Institute, which is operated by the Association of Universities for Research in Astronomy, Inc., under NASA contract NAS 5–26555. 

This publication makes use of data products from the Two Micron All Sky Survey, which is a joint project of the University of Massachusetts and the Infrared Processing and Analysis Center/California Institute of Technology, funded by the National Aeronautics and Space Administration and the National Science Foundation. 

The national facility capability for SkyMapper has been funded through ARC LIEF grant LE130100104 from the Australian Research Council, awarded to the University of Sydney, the Australian National University, Swinburne University of Technology, the University of Queensland, the University of Western Australia, the University of Melbourne, Curtin University of Technology, Monash University and the Australian Astronomical Observatory. SkyMapper is owned and operated by The Australian National University's Research School of Astronomy and Astrophysics. The survey data were processed and provided by the SkyMapper Team at ANU. The SkyMapper node of the All-Sky Virtual Observatory (ASVO) is hosted at the National Computational Infrastructure (NCI). Development and support the SkyMapper node of the ASVO has been funded in part by Astronomy Australia Limited (AAL) and the Australian Government through the Commonwealth's Education Investment Fund (EIF) and National Collaborative Research Infrastructure Strategy (NCRIS), particularly the National eResearch Collaboration Tools and Resources (NeCTAR) and the Australian National Data Service Projects (ANDS). 

Funding for the Sloan Digital Sky Survey IV has been provided by the Alfred P. Sloan Foundation, the U.S. Department of Energy Office of Science, and the Participating Institutions. SDSS acknowledges support and resources from the Center for High-Performance Computing at the University of Utah. The SDSS web site is www.sdss.org. SDSS is managed by the Astrophysical Research Consortium for the Participating Institutions of the SDSS Collaboration including the Brazilian Participation Group, the Carnegie Institution for Science, Carnegie Mellon University, Center for Astrophysics | Harvard \& Smithsonian (CfA), the Chilean Participation Group, the French Participation Group, Instituto de Astrofísica de Canarias, The Johns Hopkins University, Kavli Institute for the Physics and Mathematics of the Universe (IPMU) / University of Tokyo, the Korean Participation Group, Lawrence Berkeley National Laboratory, Leibniz Institut für Astrophysik Potsdam (AIP), Max-Planck-Institut für Astronomie (MPIA Heidelberg), Max-Planck-Institut für Astrophysik (MPA Garching), Max-Planck-Institut für Extraterrestrische Physik (MPE), National Astronomical Observatories of China, New Mexico State University, New York University, University of Notre Dame, Observatório Nacional / MCTI, The Ohio State University, Pennsylvania State University, Shanghai Astronomical Observatory, United Kingdom Participation Group, Universidad Nacional Autónoma de México, University of Arizona, University of Colorado Boulder, University of Oxford, University of Portsmouth, University of Utah, University of Virginia, University of Washington, University of Wisconsin, Vanderbilt University, and Yale University.

This research has made use of the NASA Exoplanet Archive, which is operated by the California Institute of Technology under contract with NASA under the Exoplanet Exploration Program.  

This research has made use of the SIMBAD database, operated at Centre de Donn\'ees astronomiques de Strasbourg (CDS), Strasbourg, France.

This research has made use of the VizieR catalog access tool, CDS, Strasbourg, France (DOI: \href{http://doi.org/10.26093/cds/vizier}{10.26093/cds/vizier}). The original description of the VizieR service was published in A\&AS 143, 23 \citep{SIMBAD}. 

This research made use of Astropy,\footnote{http://www.astropy.org} a community-developed core Python package for Astronomy \citep{astropy:2013, astropy:2018}. 
%\end{acknowledgments*}

\software{pandas \citep{pandas}, Astropy \citep{astropy:2013, astropy:2018}, MultiNest \citep{Feroz2008, Feroz2009, Feroz2019}, isochrones \citep{Morton2015}, pyia \citep{PriceWhelan2018}, MESA \citep{Paxton2011, Paxton2013, Paxton2015, Paxton2018, Paxton2019}, matplotlib \citep{matplotlib}, scikit-learn \citep{sklearn}}
\facilities{NASA Exoplanet Archive \citep{Akeson2013, NASAExoArcdoi}, GALEX, Skymapper, Gaia, CTIO:2MASS, WISE, Sloan}

\bibliography{mybib_v2}{}
\bibliographystyle{aasjournal}

\appendix
\label{appendix}
We apply suggested quality cuts for each catalog to ensure high quality photometry. 

\textbf{GALEX}: To ensure good GALEX NUV magnitudes, we remove any measurements with the flags 2, 4, or 128 set in \texttt{NUV\_artifact} (called \texttt{Nafl} on VizieR). These values correspond to the base 10 equivalents of bits that are set for each flag. These cuts ensure that the measurement was not impacted by the window reflection of a bright star in the NUV, a dichroic reflection, and was not masked out due to variability.\footnote{See the \href{https://galex.stsci.edu/GR6/?page=ddfaq\#6}{GALEX Data Description} or \citep{Galex} for more details on these flags.} 

\textbf{SkyMapper DR2}: For each band ``x'' measured by \texttt{SkyMapper}, we keep measurements with \texttt{flagsx}=0, \texttt{Niflagsx}=0, and \texttt{Ngoodx}$>$1. This set of flags ensures that the measurements were not affected by saturation, contamination, position on the detector, bad pixels, cross-talk, or cosmic rays and that there was at least one good measurement. \footnote{See \href{http://skymapper.anu.edu.au/data-release/}{the SkyMapper DR2 release notes} for further details.} 

\textbf{SDSS DR13}: We require that for a band ``x'', \texttt{((flags\_x \& 0x10000000) != 0)}, \texttt{((flags\_x \& 0x8100000c00a4) = 0)}, \texttt{(((flags\_x \& 0x400000000000) = 0) or (psfmagerr\_x <= 0.2))}, and that \texttt{(((flags\_x \& 0x100000000000) = 0) or (flags\_x \& 0x1000) = 0)}. These quality flags ensure that the object was detected in a 1x1 binned image, that it is not too close to the edge of the frame to estimate a radial profile, that there are no saturated pixels, that most of the flux is not from interpolated pixels, that the number of interpolated pixels does not affect the reliability of the error estimate, and that the measurement was not affected by cosmic rays\footnote{For more details, see the \href{http://skyserver.sdss.org/dr13/en/help/browser/browser.aspx?cmd=description+PhotoObjAll+U\#\&\&history=enum+PhotoFlags+E}{description of the SDSS DR13 photometry flags data values}.}

\textbf{2MASS}: We keep J, H, and K magnitudes when the \texttt{ph\_qual} is A or the \texttt{rd\_flg} is 1 or 3, the \texttt{cc\_flg} is 0, the \texttt{bl\_flg} is 1, and \texttt{use\_src} is 1. These flags ensure that the source was detected at a high SNR and that the photometry is high quality, that there is no source confusion or contamination, there was no blending, and that the best detection of the source is used.\footnote{For more details, see the \href{https://irsa.ipac.caltech.edu/data/2MASS/docs/releases/allsky/doc/sec1_6b.html}{2MASS documentation}.} 

\textbf{AllWISE}: We keep the $W1$ and $W2$ magnitudes when \texttt{cc\_flags} is one of 0, d, p, h, or o, and \texttt{ext\_flg} is 0. These quality flags ensure that there the measurement is not a spurious detection of an artifact and that there is not extended source contamination. \footnote{See the \href{https://wise2.ipac.caltech.edu/docs/release/allwise/expsup/sec2_1a.html}{AllWISE Source Catalog and Reject Table} for more details.}

\startlongtable
\begin{deluxetable}{ccc}
\tabletypesize{\footnotesize}
\tablecaption{Discovery \& Obliquity References}
\tablenum{6}
\tablewidth{0pt}
\tablehead{\colhead{Planet} & \colhead{Discovery Reference} & \colhead{Obliquity Reference}}
\startdata
WASP-32 b & \citet{Maxted2010b} & \citet{Brown2012}\\
WASP-26 b & \citet{Smalley2010} & \citet{Albrecht2012}\\
WASP-1 b & \citet{CollierCameron2007} & \citet{Albrecht2011}\\
WASP-190 b & \citet{Temple2019} & \citet{Temple2019}\\
HAT-P-16 b & \citet{Buchhave2010} & \citet{Albrecht2012}\\
WASP-18 b & \citet{Hellier2009b} & \citet{Triaud2010}\\
WASP-76 b & \citet{West2016} & \citet{Ehrenreich2020}\\
WASP-71 b & \citet{Smith2013} & \citet{Brown2017}\\
WASP-53 b & \citet{Triaud2017} & \citet{Triaud2017}\\
WASP-72 b & \citet{Gillon2013} & \citet{Addison2018}\\
WASP-11 b & \citet{West2009} & \citet{Mancini2015}\\
WASP-22 b & \citet{Maxted2010a} & \citet{Anderson2011}\\
K2-29 b & \citet{Santerne2016} & \citet{Santerne2016}\\
WASP-78 b & \citet{Smalley2012} & \citet{Brown2017}\\
WASP-79 b & \citet{Smalley2012} & \citet{Johnson2017}\\
WASP-61 b & \citet{Hellier2012} & \citet{Brown2017}\\
KELT-7 b & \citet{Bieryla2015} & \citet{Zhou2016}\\
EPIC 246851721 b & \citet{Yu2018} & \citet{Yu2018}\\
WASP-62 b & \citet{Hellier2012} & \citet{Brown2017}\\
WASP-49 b & \citet{Lendl2012} & \citet{Wyttenbach2017}\\
XO-6 b & \citet{Crouzet2017} & \citet{Crouzet2017}\\
WASP-12 b & \citet{Hebb2009} & \citet{Albrecht2012}\\
CoRoT-18 b & \citet{Hebrard2011} & \citet{Hebrard2011}\\
HAT-P-56 b & \citet{Huang2015} & \citet{Zhou2016}\\
WASP-121 b & \citet{Delrez2016} & \citet{Bourrier2020}\\
HAT-P-24 b & \citet{Kipping2010} & \citet{Albrecht2012}\\
HAT-P-9 b & \citet{Shporer2009} & \citet{Moutou2011}\\
XO-4 b & \citet{McCullough2008} & \citet{Narita2010}\\
KELT-19 A b & \citet{Siverd2018} & \citet{Siverd2018}\\
HAT-P-20 b & \citet{Bakos2011} & \citet{Esposito2017}\\
XO-2 N b & \citet{Burke2007} & \citet{Damasso2015}\\
HAT-P-30 b & \citet{Johnson2011} & \citet{Johnson2011}\\
KELT-17 b & \citet{Zhou2016} & \citet{Zhou2016}\\
K2-34 b & \citet{Hirano2016} & \citet{Hirano2016}\\
HAT-P-13 b & \citet{Bakos2009} & \citet{Winn2010}\\
HAT-P-69 b & \citet{Zhou2019} & \citet{Zhou2019}\\
WASP-84 b & \citet{Anderson2014} & \citet{Anderson2015}\\
WASP-13 b & \citet{Skillen2009} & \citet{Brothwell2014}\\
WASP-166 b & \citet{Hellier2019} & \citet{Hellier2019}\\
MASCARA-4 b & \citet{Dorval2020} & \citet{Ahlers2020}\\
WASP-19 b & \citet{Hebb2010} & \citet{Sedaghati2021}\\
WASP-43 b & \citet{Hellier2011} & \citet{Esposito2017}\\
HAT-P-22 b & \citet{Bakos2011} & \citet{Mancini2018}\\
WASP-66 b & \citet{Hellier2012} & \citet{Addison2016}\\
WASP-127 b & \citet{Lam2017} & \citet{Allart2020}\\
KELT-24 b & \citet{Rodriguez2019} & \citet{Rodriguez2019}\\
WASP-31 b & \citet{Anderson2011} & \citet{Brown2012}\\
WASP-87 b & \citet{Addison2016} & \citet{Addison2016}\\
K2-140 b & \citet{Giles2018} & \citet{Rice2021}\\
HAT-P-36 b & \citet{Bakos2012} & \citet{Mancini2015}\\
WASP-41 b & \citet{Maxted2011} & \citet{Southworth2016}\\
WASP-25 b & \citet{Enoch2011} & \citet{Brown2012}\\
KELT-6 b & \citet{Collins2014} & \citet{Damasso2015}\\
WASP-167 b & \citet{Temple2017} & \citet{Temple2017}\\
HAT-P-3 b & \citet{Torres2007} & \citet{Mancini2018}\\
Qatar-2 b & \citet{Bryan2012} & \citet{Mocnik2017}\\
WASP-15 b & \citet{West2009} & \citet{Triaud2010}\\
HAT-P-12 b & \citet{Hartman2009} & \citet{Mancini2018}\\
NGTS-2 b & \citet{Raynard2018} & \citet{Anderson2018}\\
WASP-39 b & \citet{Faedi2011} & \citet{Mancini2018}\\
WASP-14 b & \citet{Joshi2009} & \citet{Johnson2009}\\
WASP-24 b & \citet{Street2010} & \citet{Simpson2011}\\
HAT-P-4 b & \citet{Kovacs2007} & \citet{Winn2011}\\
WASP-17 b & \citet{Anderson2010} & \citet{Triaud2010}\\
WASP-38 b & \citet{Barros2011} & \citet{Brown2012}\\
HAT-P-2 b & \citet{Bakos2007b} & \citet{Loeillet2008}\\
HD 149026 b & \citet{Sato2005} & \citet{Albrecht2012}\\
WASP-103 b & \citet{Gillon2014} & \citet{Addison2016}\\
TOI-2109 b & \citet{Wong2021} & \citet{Wong2021}\\
WASP-148 b & \citet{Hebrard2020} & \citet{Wang2021}\\
HAT-P-18 b & \citet{Hartman2011} & \citet{Esposito2014}\\
HAT-P-14 b & \citet{Torres2010} & \citet{Winn2011}\\
WASP-3 b & \citet{Pollacco2008} & \citet{Miller2010}\\
Kepler-8 b & \citet{Jenkins2010} & \citet{Albrecht2012}\\
CoRoT-2 b & \citet{Alonso2008} & \citet{Bouchy2008}\\
HAT-P-7 b & \citet{Pal2008} & \citet{Benomar2014}\\
KELT-20 b & \citet{Lund2017} & \citet{Lund2017}\\
HAT-P-41 b & \citet{Hartman2012} & \citet{Johnson2017}\\
Kepler-17 b & \citet{Desert2011} & \citet{Desert2011}\\
HD 189733 b & \citet{Bouchy2005} & \citet{Cegla2016}\\
WASP-80 b & \citet{Triaud2013} & \citet{Triaud2015}\\
HAT-P-34 b & \citet{Bakos2012} & \citet{Albrecht2012}\\
WASP-74 b & \citet{Hellier2015} & \citet{Luque2020}\\
KELT-21 b & \citet{Johnson2018} & \citet{Johnson2017}\\
HAT-P-23 b & \citet{Bakos2011} & \citet{Moutou2011}\\
KELT-9 b & \citet{Gaudi2017} & \citet{Wyttenbach2020}\\
WASP-7 b & \citet{Hellier2009a} & \citet{Albrecht2011}\\
WASP-94 A b & \citet{Neveu-VanMalle2014} & \citet{Neveu-VanMalle2014}\\
WASP-69 b & \citet{Anderson2014} & \citet{Casasayas-Barris2017}\\
HD 209458 b & \citet{Henry2000} & \citet{Casasayas-Barris2021}\\
WASP-47 b & \citet{Hellier2012} & \citet{Sanchis-Ojeda2015}\\
HAT-P-8 b & \citet{Latham2009} & \citet{Moutou2011}\\
HAT-P-1 b & \citet{Bakos2007a} & \citet{Johnson2008}\\
WASP-21 b & \citet{Bouchy2010} & \citet{Chen2020b}\\
WASP-6 b & \citet{Tregloan-Reed2015} & \citet{Tregloan-Reed2015}\\
WASP-52 b & \citet{Hebrard2013} & \citet{Chen2020a}\\
TOI-1518 b & \citet{Cabot2021} & \citet{Cabot2021}\\
WASP-4 b & \citet{Wilson2008} & \citet{Triaud2010}\\
WASP-28 b & \citet{Petrucci2015} & \citet{Anderson2015}\\
HAT-P-6 b & \citet{Noyes2008} & \citet{Albrecht2012}\\
WASP-60 b & \citet{Hebrard2013} & \citet{Mancini2018}\\
WASP-5 b & \citet{Anderson2008} & \citet{Triaud2010}\\
\enddata
\end{deluxetable}

\end{document}